\documentclass[reprint,amsmath,amssymb,aps,superscriptaddress]{revtex4-2}

\usepackage{graphicx}
\usepackage{dcolumn}
\usepackage{bm}
\usepackage{physics}
\usepackage{adjustbox}

\begin{document}

\preprint{APS/123-QED}

\title{Double-helicoid surface states in Dirac semimetals protected by glide-time-reversal symmetry}

\author{Taiki Yukitake}
\affiliation{Department of Physics, Institute of Science Tokyo, 2-12-1 Ookayama,
Meguro-ku, Tokyo 152-8551, Japan}
\affiliation{Department of Physics, Tokyo Institute of Technology, 2-12-1 Ookayama, Meguro-ku, Tokyo 152-8551, Japan}
\author{Daisuke Hara}
\affiliation{Department of Physics, Tokyo Institute of Technology, 2-12-1 Ookayama, Meguro-ku, Tokyo 152-8551, Japan}
\author{Shuichi Murakami}
\email{murakami@stat.phys.titech.ac.jp}
\affiliation{Department of Physics, Institute of Science Tokyo, 2-12-1 Ookayama,
Meguro-ku, Tokyo 152-8551, Japan}
\affiliation{Department of Physics, Tokyo Institute of Technology, 2-12-1 Ookayama, Meguro-ku, Tokyo 152-8551, Japan}
\affiliation{International Institute for Sustainability with Knotted Chiral Meta
Matter (WPI-SKCM$^2$), Hiroshima University, Higashi-hiroshima, Hiroshima
739-0046, Japan}

\date{\today}

\begin{abstract}
Recently, some $Z_2$ monopole charges were defined for Dirac semimetals with $\mathcal{GT}$ symmetry ($\mathcal{G}$: glide, $\mathcal{T}$: time-reversal) in previous works, and the charges are believed to lead to double-helicoid surface states. However, no proof of the bulk-surface correspondence is given there. In this paper, we point out one of the $Z_2$ charges in the previous works is gauge-dependent, and newly define another $Z_2$ charge. Using this new $Z_2$ charge, we give a proof of the bulk-surface correspondence. We also compare the new $Z_2$ charge with the $Z_2$ invariant for $\mathcal{G}$-protected topological crystalline insulators, and the second Stiefel-Whitney number for $\mathcal{PT}$-protected nodal line semimetals.
\end{abstract}

\maketitle

\section{\label{sec:1}Introduction}
The study of topological semimetals (SMs) is one of the main topics in modern condensed matter physics. After extensive searches for topological SMs, many types of them are now recognized, such as Weyl SMs~\cite{Murakami2007, Wan2011, Li2015, Xu2015, Lv2015} and Dirac SMs~\cite{Young2012, Yang2014, Yang2015, Gao2016, Wang2012, Liu2014, Wang2013, Borisenko2014, Morimoto2014, Tang2016, Lin2020, Qian2023}. To understand topological SMs, it is important to study protection mechanisms of gapless nodes in the bulk Brillouin zone (BZ) and the bulk-surface correspondence for the gapless nodes. For example, Weyl points in three-dimensional (3D) Weyl SMs are associated with the monopole charge $C(=\pm 1)$, which is defined as the Chern number on a sphere enclosing the Weyl points~\cite{Wan2011}. It protects the existence of the Weyl points against perturbations. Moreover, it leads to helicoid surface states (HSSs) around the projection of the Weyl points on the surface BZ, whose dispersion along a loop enclosing the projection of the Weyl points is chiral~\cite{Wan2011, Li2015}. Meanwhile, Dirac points in 3D Dirac SMs are not stable against perturbations in general, and no topological gapless surface states appear around the projection of the Dirac points because the monopole charge $C$ for the Dirac points is equal to zero. Researchers have studied additional symmetry conditions that force the Dirac points to have nontrivial surface states. 

In this context, Dirac SMs with the composition of glide symmetry ($\mathcal{G}$) and time-reversal symmetry ($\mathcal{T}$) were proposed~\cite{Fang2016, Zhang2022, Hara2023, Zhang2023, Cheng2020, Cai2020, Su2022}. In the $\mathcal{GT}$-protected Dirac SMs, some $Z_2$ charges have been defined, and they are believed to lead to double-helicoid surface states (DHSSs) around the projection of Dirac points on the surface BZ. Here, DHSSs are surface states whose dispersion along a loop enclosing the projection of the Dirac points is helical, and they can be regarded as the superposition of HSSs and anti-HSSs.
However, no proof of the bulk-surface correspondence is given there. Moreover, we found that the discussions on the gauge-independency of the $Z_2$ charge in Ref.~\cite{Zhang2022} have some problems.

In this paper, we point out that the $Z_2$ charge defined in Ref.~\cite{Zhang2022} is ill-defined, and define another $Z_2$ charge. Then we can give a proof of the bulk-surface correspondence for the new $Z_2$ charge. We also show some important properties of the new $Z_2$ charge such as the relationship with the $Z_2$ invariant for $\mathcal{G}$-protected topological crystalline insulators~\cite{Shiozaki2015, Fang2015, Kim2019, Kim2020, Zhang2020, Zhou2021}, and the second Stiefel-Whitner (SW) number for $\mathcal{PT}$-protected $Z_2$ nodal line SMs~\cite{Fang2015-2, Zhao2017, Ahn2018, Wang2020, Chen2022, Xue2023, Xiang2024, Ma2024, Wang2024, Yue2024} . Moreover, we develop computation methods for the new $Z_2$ charge. This paper is organized as follows. In Sec.~\ref{sec:2}, we review previous studies of $Z_2$ charges in $\mathcal{GT}$-protected Dirac SMs. In Sec.~\ref{sec:3}, we point out the ill-definedness of one of the previous $Z_2$ charges and define another $Z_2$ charge. In Sec.~\ref{sec:4}, we prove the bulk-surface correspondence for the new $Z_2$ charge. In Sec.~\ref{sec:5}, we compare the new $Z_2$ charge with other topological invariants under some additional symmetries. In Sec.~\ref{sec:6}, we explain formulas of the new $Z_2$ charge useful for computation, using Wilson loop methods or Fu-Kane like formulas. In Sec.~\ref{sec:7}, we give some tight-binding models with the $\mathcal{GT}$ symmetry and examine our discussion. We conclude this paper in Sec.~\ref{sec:8}.

Below, we mainly consider Dirac SMs with $\mathcal{GT}$ symmetry. Meanwhile, the discussions below also hold for other topological SMs with the $\mathcal{GT}$ symmetry, which are realized by perturbing the above mentioned Dirac SMs. These topological SMs include Weyl SMs with Weyl dipoles (pairs of Weyl points related by the $\mathcal{GT}$ symmetry) and nodal line SMs with $Z_2$ nodal rings. Also, although we focus on systems on a primitive lattice, which corresponds to the magnetic space group (MSG) \#7.26 ($Pc'$), almost the same discussions hold for systems on a non-primitive lattice, which corresponds to MSG \#9.39 ($Cc'$).

\begin{figure}
\includegraphics[width=\columnwidth]{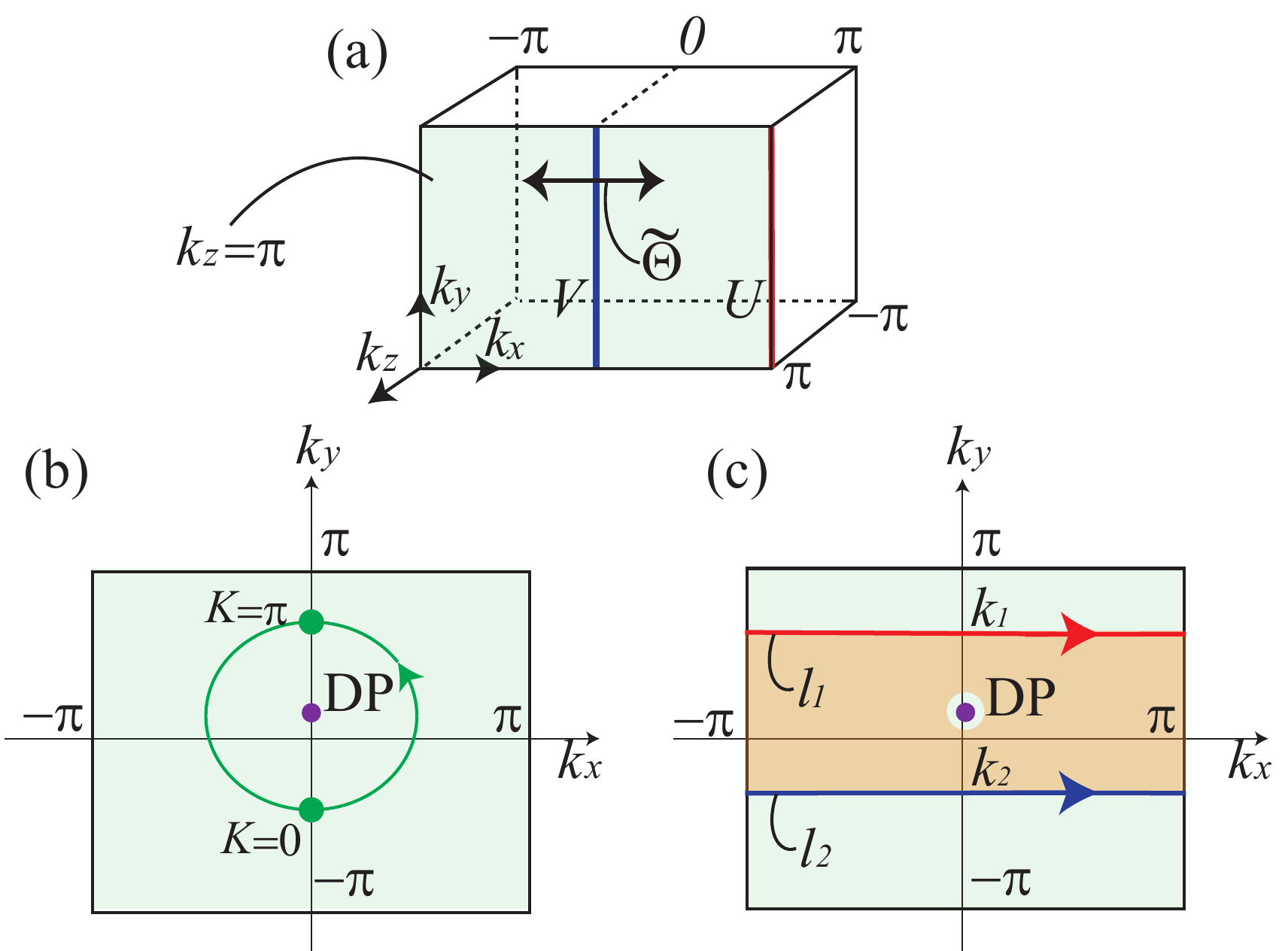}
\caption{\label{fig1}\small{Previous $Z_2$ charges in Dirac SMs with $\tilde{\Theta}=\mathcal{GT}$ symmetry. (a) Bulk BZ for MSG \#7.26 ($Pc'$). $\tilde{\Theta}^2=-1$ holds on the plane $k_z=\pi$. Points on the lines $U$ and $V$ are $\tilde{\Theta}$-invariant.  (b) $Z_2$ charge $\mathcal{Q}$. A circle on the plane $k_z=\pi$ is used to define $\mathcal{Q}$. (c) $Z_2$ charge $\tilde{\mathcal{Q}}$. Two lines $l_1$ and $l_2$ on the plane $k_z=\pi$ are used to define $\tilde{\mathcal{Q}}$.}}
\end{figure}

\section{\label{sec:2}Previous $Z_2$ charges in Dirac semimetals with $\mathcal{GT}$ symmetry} 

In this section, we review the studies of $Z_2$ charges in Dirac SMs with $\mathcal{GT}$ symmetry in Refs.~\cite{Fang2016, Zhang2022}. We consider Dirac SMs with MSG \#7.26 ($Pc'$). This MSG is generated by $\tilde{\Theta}=\mathcal{GT}=\{M_y|0 0 \frac{1}{2}\}'$ and translation operations $\{E |100\}, \{E |010\}, \{E |001\}$, where $M_y$ represents the mirror reflection with respect to the $x$-$z$ plane, $'$ represents the time-reversal operation, and $E$ represents the identity operation. Here, we set the lattice constants to be unity for simplicity. 
In such Dirac SMs, the Bloch Hamiltonian $H(\bm{k})$ and the operator $\tilde{\Theta}(\bm{k})$ which represents the operation $\tilde{\Theta}$ under the Bloch basis set must satisfy 
\begin{align}
\tilde{\Theta}(\bm{k})H(\bm{k})\tilde{\Theta}(\bm{k})^{-1} &=H(\bm{k}'), \label{eq:2-0} \\
\tilde{\Theta}(\bm{k}')\tilde{\Theta}(\bm{k}) &=e^{-ik_z},  \label{eq:2-1}
\end{align}
where $\bm{k}'=\tilde{\Theta}\bm{k}=(-k_x, k_y, -k_z)$. From Eq.~(\ref{eq:2-1}), we have $\tilde{\Theta}^2=-1$ on the plane $k_z=\pi$. As a consequence, the sewing matrix $\omega(\bm{k})$ becomes antisymmetric at $\tilde{\Theta}$-invariant points on the plane $k_z=\pi$ (see Fig.~\ref{fig1}(a)), where $\omega(\bm{k})$ is defined as 
$[\omega (\bm{k})]_{mn}=\mel{u_{m\tilde{\Theta}\bm{k} }}{\tilde{\Theta}}{u_{n\bm{k}}}\ (m,n=1,2,\dots N_{\mathrm{occ}}/2)$, $\ket{u_{n\bm{k}}}$ is the periodic part of the $n$-th occupied Bloch band, and $N_{\mathrm{occ}}$ is the total number of occupied bands. 

For the Dirac SMs, a $Z_2$ charge is defined in Ref.~\cite{Fang2016}. Let us consider a sphere whose center is on the lines $U$ or $V$, and take a smooth gauge over the sphere. Then the $Z_2$ charge $\mathcal{Q}$ is defined as
\begin{align}
(-1)^{\mathcal{Q}}=\frac{\mathrm{Pf}[\omega(K=0)]}{\sqrt{\mathrm{det}[\omega(K=0)]}}\frac{\mathrm{Pf}[\omega(K=\pi)]}{\sqrt{\mathrm{det}[\omega(K=\pi)]}}, \label{eq:2-3}
\end{align}
where $K\in[-\pi,\pi]$ parameterizes the cross section of the sphere by the plane $k_z=\pi$ so that $\tilde{\Theta}$ transforms $K$ to $-K$ as shown in Fig.~\ref{fig1}(b).
It is claimed that the value of $\mathcal{Q}$ becomes nontrivial when the sphere encloses a Dirac point, and the nontrivial $\mathcal{Q}$ leads to DHSSs, but without giving their proofs.  

Meanwhile, another $Z_2$ charge is defined in Ref.~\cite{Zhang2022}. Let us consider two lines $l_i\ (i=1,2)$ parallel to $k_x$ axis in the bulk BZ shown as red and blue lines in Fig.~\ref{fig1}(c), and take a smooth gauge over the plane $\{\bm{k}| -\pi\le k_x\le \pi,\ k_2\le k_y\le k_1,\ k_z=\pi\}$ except for Dirac points on the plane. Because the lines are mapped onto themselves by $\tilde{\Theta}$, on each line, we can divide the set of $N_{\mathrm{occ}}$ occupied bands into two groups $\alpha$ and $\beta$ so that $\tilde{\Theta}$ transforms the two groups mutually; when $\ket{u_{n\bm{k}}^{(\alpha)}}\ (n=1,2,\dots, N_{\mathrm{occ}}/2)$ denote the occupied states belong to the group $\alpha$, then $\tilde{\Theta}\ket{u_{n\bm{k}}^{(\alpha)}}$ are in the subspace spanned by $\ket{u_{m\tilde{\Theta}\bm{k}}^{(\beta)}}\ (m=1,2,\dots, N_{\mathrm{occ}}/2)$, and vice versa. Under the situation, we can define the Berry connection and the Berry phase along the lines $l_i\ (i=1,2)$ for the groups $j=\alpha,\ \beta$ as
\begin{align}
\bm{A}^{(j)}(\bm{k}) &=\sum_{n=1}^{N_{\mathrm{occ}}/2}i\mel{u_{n\bm{k}}^{(j)}}{\nabla_{\bm{k}}}{u_{n\bm{k}}^{(j)}}, \label{eq:2-4}\\
\gamma^{(j)} [l_i] &=\int_{l_i}d\bm{k}\cdot\bm{A}^{(j)}(\bm{k}).  \label{eq:2-5}
\end{align}
They are the $\alpha$ and $\beta$ parts of the Berry connection $\bm{A}(\bm{k})$ and the Berry phase $\gamma[l_i]$: $\bm{A}(\bm{k})=\bm{A}^{(\alpha)}(\bm{k})+\bm{A}^{(\beta)}(\bm{k})$, $\gamma[l_i]=\gamma^{(\alpha)} [l_i]+\gamma^{(\beta)} [l_i]$. Then, the $\tilde{\Theta}$-polarization $P_{\tilde{\Theta}}[l_i]$ on the lines $l_i\ (i=1,2)$ is defined as 
\begin{align}
P_{\tilde{\Theta}}[l_i]=\frac{1}{2\pi}\qty(\gamma^{(\alpha)}[l_i]-\gamma^{(\beta)}[l_i]). \label{eq:2-6}
\end{align}
Finally, the $Z_2$ charge $\tilde{\mathcal{Q}}$ is defined as the difference of the $\tilde{\Theta}$-polarization on the two lines:
\begin{align}
\tilde{\mathcal{Q}}=P_{\tilde{\Theta}}[l_1]-P_{\tilde{\Theta}}[l_2] \ \ (\mathrm{mod}\ 2).  \label{eq:2-7}
\end{align}
It is claimed that the value of $\tilde{\mathcal{Q}}$ becomes nontrivial when a Dirac point lies between the two lines $l_1$ and $l_2$, and the nontrivial $\tilde{\mathcal{Q}}$ leads to DHSSs. A proof of the bulk-surface correspondence is given in Ref.~\cite{Zhang2022}, but the proof is valid only when Dirac systems are spinless and have both $\mathcal{G}$ and $\mathcal{T}$ symmetries. 

Here, in Ref.~\cite{Zhang2022}, it is claimed that $\mathcal{Q}$ is gauge-dependent and $\tilde{\mathcal{Q}}$ is gauge-independent. However, we find that this claim is incorrect: actually, $\mathcal{Q}$ is gauge-independent but $\tilde{\mathcal{Q}}$ is gauge-dependent. We explain this consequence in the next section. Furthermore, we also show that, by slightly modifying the definition of $\tilde{\mathcal{Q}}$, we can define a new $Z_2$ charge, which is gauge-independent and reduces to $\mathcal{Q}$ in a special case.

\begin{figure}
\includegraphics[width=\columnwidth]{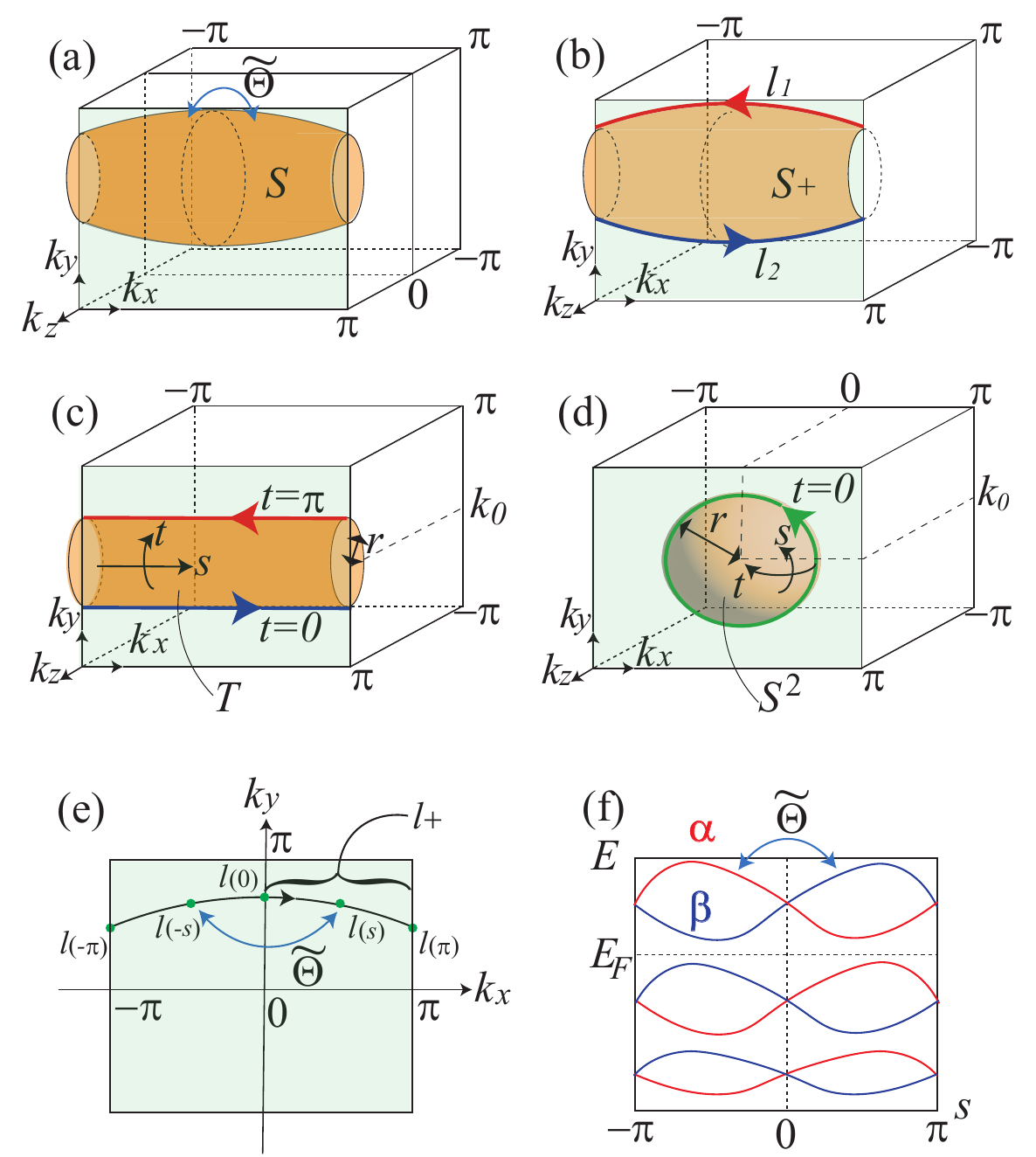}
\caption{\label{fig2}\small{New $Z_2$ charge $\mathcal{Q}[S]$. (a) Closed surface $S$ in the bulk BZ used to define $\mathcal{Q}[S]$. $S$ is closed under $\tilde{\Theta}$ and not intersect with the plane $k_z=0$. (b) Surface $S_+$. It is defined as a half of the surface $S$ with $k_z\ge\pi$, and thus it has a boundary, e.g. $\partial S_+=l_1\cup l_2$ in the figure. (c,d) Examples of $S$ satisfying the conditions. (c) Torus $T$ defined as $T=\{\bm{k} | (k_x, k_y, k_z)=(s, k_0-r\cos t, \pi+r\sin t)  \ \ (-\pi\le s\le\pi\ ,-\pi\le t\le\pi)\}$, (d) Sphere $S^2$ defined as $S^2=\{\bm{k}| (k_x, k_y, k_z)=(r\cos\frac{t}{2}\cos s, k_0+r\cos\frac{t}{2}\sin s, \pi+r\sin\frac{t}{2})\ \ (-\pi\le s\le\pi\ ,-\pi\le t\le\pi)\}$. Here, $r(>0)$ and $k_0$ are constants. (e) Closed line $l$ used to define $P[l]$. It is closed under $\tilde{\Theta}$ and parameterized by $s\in [-\pi,\pi]$ so that $\tilde{\Theta}l(s)=l(-s)$. (f) Band structure on the line $l$ in (e). The bands can be divided into two groups $\alpha$ and $\beta$ so that $\tilde{\Theta}$ transforms the two groups mutually. }}
\end{figure}

\section{\label{sec:3}Redefinition of $Z_2$ Charge}
In this section, we point out that the previous $Z_2$ charge $\tilde{\mathcal{Q}}$ is ill-defined, and newly define another $Z_2$ charge $\mathcal{Q}[S]$, which can be defined for a closed surface $S$ in the bulk BZ satisfying given conditions. As we observe below, this new $Z_2$ charge can be seen as a modification of $\tilde{\mathcal{Q}}$ and reduces to $\mathcal{Q}$, with proper $S$s. Therefore, we can unify the discussions on the $Z_2$ charges in Refs.~\cite{Fang2016, Zhang2022} (see Secs.~\ref{sec:5-1} and \ref{sec:6}). Moreover, we can derive some new consequences which cannot be obtained from the previous $Z_2$ charges by using $\mathcal{Q}[S]$ with proper $S$s (see Secs.~\ref{sec:3-2}, \ref{sec:4}, and \ref{sec:5-2}).

\subsection{\label{sec:3-0}Ill-definedness of the previous $Z_2$ charge $\tilde{\mathcal{Q}}$}
In the previous section, we reviewed the definition of the $Z_2$ charges $\mathcal{Q}$ and $\tilde{\mathcal{Q}}$ for Dirac SMs with $\tilde{\Theta}$. However, in the present paper, we find that $\tilde{\mathcal{Q}}$ is ill-defined. Actually, the gauge condition for
$\tilde{\mathcal{Q}}$ permits the existence of a gauge transformation with singularity at gapless points, which alters the value of $\tilde{\mathcal{Q}}$ by unity. We give an example of such gauge transformations in App.~\ref{sec:A}. Meanwhile, we find $\mathcal{Q}$ is gauge-independent by comparing it with the newly defined $Z_2$ charge $\mathcal{Q}[S]$ (see Sec.~\ref{sec:3-2-1}). This answers the question on the gauge-independency of $\mathcal{Q}$ doubted in Ref.~\cite{Zhang2022}.

\subsection{\label{sec:3-1}Definition of the new $Z_2$ charge $\mathcal{Q}[S]$}
We define the new $Z_2$ charge $\mathcal{Q}[S]$ for an oriented closed surface $S$ satisfying the conditions:
\begin{enumerate}
    \item $S$ is closed under $\tilde{\Theta}: (k_x, k_y, k_z)\mapsto (-k_x, k_y, -k_z)$,
    \item $S$ does not intersect with the plane $k_z=0$,
    \item the system is gapped over $S$,  
\end{enumerate}
as shown in Fig.~\ref{fig2}(a). This surface $S$ is divided into half by the plane $k_z=\pi$. Then, $S_+$ denotes the divided part with $k_z\ge\pi$, and $\partial S_+$ denotes the boundary of $S_+$ on the plane $k_z=\pi$, as shown in Fig.~\ref{fig2}(b). A torus $T$ shown in Fig.~\ref{fig2}(c) and a sphere $S^2$ shown in Fig.~\ref{fig2}(d) are examples of such surface $S$.

The definition of $\mathcal{Q}[S]$ is as follows:
\begin{align}
    \mathcal{Q}[S]=\frac{1}{2\pi}\int_{S_+}d\bm{S}\cdot\mathrm{rot}\bm{A}(\bm{k})-2\sum_{l\in\partial S_+}P[l]\ \ \ (\mathrm{mod}\ 2), \label{eq:3-1}
\end{align}
where $l\in\partial S_+$ means that the curve $l$ is one of the connected components of the boundary of $S_+$, and $P[l]$ is defined as
\begin{align}
P[l]=\frac{1}{2\pi}\qty[\int_{l_+}d\bm{k}\cdot\bm{A}(\bm{k})+i\log\qty(\frac{\mathrm{Pf}\omega [l(\pi)]}{\mathrm{Pf}\omega [l(0)]})]\ (\mathrm{mod}\ 1) \label{eq:3-2}
\end{align}
for a curve $l$ closed under $\tilde{\Theta}$ on the plane $k_z=\pi$. Here, $s\in[-\pi,\pi]$ parameterizes $l$ so that $\tilde{\Theta}l(s)=l(-s)$, and $l_+$ is a half of $l$ with $0\le s \le \pi$, as shown in Fig.~\ref{fig2}(e). The quantity $P[l]$ can be related with the Berry phase on the line $l$ as
\begin{align}
    P[l]=\frac{1}{2\pi}\gamma^{(\alpha)}[l] \ \ (\mathrm{mod}\ 1), \label{eq:3-3}
\end{align}
when we divide the set of occupied bands into two groups $\alpha$ and $\beta$, as we have presented in Sec.~\ref{sec:2} (see Fig.~\ref{fig2}(f)). We give a proof of Eq.~(\ref{eq:3-3}) in App.~\ref{sec:B}. 

$\mathcal{Q}[S]$ is gauge-independent modulo 2 because the Berry curvature is gauge-independent and $P[l]$ is gauge-independent modulo 1 (see App.~\ref{sec:B}). Moreover, the value of $\mathcal{Q}[S]$ is actually an integer because we have
\begin{align}
    (-1)^{\mathcal{Q} [S]}=\prod_{l\in\partial S_h}\frac{\mathrm{Pf}\omega [l(0)]}{\sqrt{\mathrm{det}\omega [l(0)]}}\frac{\mathrm{Pf}\omega [l(\pi)]}{\sqrt{\mathrm{det}\omega [l(\pi)]}}, \label{eq:3-4}
\end{align}
when we take a smooth gauge over $S$ (see App.~\ref{sec:B}). To summarize, $\mathcal{Q}[S]$ is a well-defined integer in terms of modulo 2. This fact also means that the value of $\mathcal{Q}[S]$ does not change under a continuous deformation of $S$ unless $S$ passes through gapless nodes under the deformation.

\subsection{\label{sec:3-2}Basic properties of the new $Z_2$ charge $\mathcal{Q}[S]$}
Before moving on to the next section, we discuss some basic properties of the new $Z_2$ charge $\mathcal{Q}[S]$, which can be obtained easily from the definition. 

\subsubsection{\label{sec:3-2-1}Relationship with the previous $Z_2$ charges $\tilde{\mathcal{Q}}$ and $\mathcal{Q}$}
First, we show that the new $Z_2$ charge $\mathcal{Q}[S]$ can be seen as a modification of the ill-defined previous $Z_2$ charge $\tilde{\mathcal{Q}}$. To this end, we take a torus $T$ defined as $T=\{\bm{k} | (k_x, k_y, k_z)=(s, k_0-r\cos t, \pi+r\sin t)  \ \ (-\pi\le s\le\pi\ ,-\pi\le t\le\pi)\}$ with constants $r(>0)$ and $k_0$ (see Fig.~\ref{fig2}(c)). Then, we have
\begin{align}
    \mathcal{Q}[T]=P_{\tilde{\Theta}}[t=\pi]-P_{\tilde{\Theta}}[t=0]\ \ (\mathrm{mod}\ 2), \label{eq:3-5}
\end{align}
from Eq.~(\ref{eq:3-3}). Equation (\ref{eq:3-5}) is almost the same as the definition of $\tilde{\mathcal{Q}}$ in Eq.~(\ref{eq:2-5}) except for the gauge condition; the gauge for $\mathcal{Q}[T]$ is taken on the torus $T$, where the system is assumed to be gapped, but that for $\tilde{\mathcal{Q}}$ is taken on the plane $k_z=\pi$, where the system is gapless at the Dirac point. As pointed out in the present paper (see App.~\ref{sec:A}), the latter gauge choice turns out to be ill-defined. It indicates that, when we consider the difference of the $\tilde{\Theta}$-polarization between the two lines in the plane $k_z=\pi$, we must choose a continuous gauge on a surface that does not pass through gapless nodes on the plane $k_z=\pi$, to make the quantity well-defined. The definition of $\mathcal{Q}[T]$ satisfies the condition, and thus we can see $\mathcal{Q}[S]$ as a modification of $\tilde{\mathcal{Q}}$

Next, we show that $\mathcal{Q}[S]$ reduces to $\mathcal{Q}$ with a proper $S$, and thus $\mathcal{Q}$ is well-defined. To this end, we take a sphere whose center is at a $\tilde{\Theta}$-invariant point, which is defined as $S^2=\{\bm{k}| (k_x, k_y, k_z)=(r\cos\frac{t}{2}\cos s, k_0+r\cos\frac{t}{2}\sin s, \pi+r\sin\frac{t}{2})\ \ (-\pi\le s\le\pi,\  -\pi\le t\le\pi)\}$ with constants $r(>0)$ and $k_0$ (see Fig.~\ref{fig2}(d)). Then, we have
\begin{align}
\mathcal{Q}[S^2]=\mathcal{Q}\ \ \ (\mathrm{mod}\ 2), \label{eq:3-6}
\end{align}
from Eq.~(\ref{eq:3-4}). Therefore, $\mathcal{Q}$ is a well-defined quantity modulo 2. In Ref.~\cite{Zhang2022}, it was claimed that $\mathcal{Q}$ is ill-defined because there is a gauge transformation which alters the value of $\mathcal{Q}$ by unity. However, this claim turns out to be incorrect. In fact, the gauge transformation mentioned in Ref.~\cite{Zhang2022} does not satisfy the gauge condition for $\mathcal{Q}$ which enforces the gauge to be smooth over the sphere. To summarize, the $Z_2$ charge $\mathcal{Q}$ defined in Ref.~\cite{Fang2016} is a special case of $\mathcal{Q}[S]$ with $S$ taken as a sphere, and therefore is a gauge-independent integer modulo 2.

\subsubsection{\label{sec:3-2-2} The value of $\mathcal{Q}[S]$ for Dirac points}
We show that a Dirac point on the $\tilde{\Theta}$-invariant lines $U$ or $V$ is $Z_2$-charged, i.e., a Dirac point has $\mathcal{Q}=1$. This is already claimed in Ref.~\cite{Fang2016}, but without its proof.
Thus, we give a proof here.

To this end, firstly, we take a sphere $S^2$ with suitable values of $k_0$ and $r$, so that the sphere encloses the Dirac point. Next, we transform the Dirac point into a pair of Weyl points located away from the plane $k_z=\pi$ by adding a $\tilde{\Theta}$-preserving perturbation. Next, we shrink $S^2$ into a point on the plane $k_z=\pi$. Then, the value of $\mathcal{Q}[S^2]$ changes by unity because $S_+$ passes through a Weyl point, and the Weyl point has the monopole charge $C=\frac{1}{2\pi}\int_{S'}d\bm{S}\cdot\mathrm{rot}\bm{A}(\bm{k})=\pm1$ for a surface $S'$ enclosing it. Meanwhile, we have $\mathcal{Q}[S^2]=0$ when $r=0$. Thus, we finally find $\mathcal{Q}=1$ for a Dirac point, with the sphere enclosing the Dirac point.

\subsubsection{\label{sec:3-2-3}Failure of Nielson-Ninomiya-like theorem}
Next, we show that the Nielson-Ninomiya (NN)-like theorem is not valid for the $Z_2$ charge $\mathcal{Q}$, i.e., the number of $Z_2$-charged gapless nodes, such as Dirac points and Weyl dipoles, in the whole BZ is not necessarily even, as opposed to the observation in Ref.~\cite{Fang2016}. 

To this end, we firstly take the boundary of the bulk BZ defined as $S_{\mathrm{tot}}=\{\bm{k} | k_x=-\pi,\pi\ \mathrm{or}\  k_y=-\pi,\pi\ \mathrm{or}\ k_z=0,2\pi\}$. The value of $\mathcal{Q}[S_{\mathrm{tot}}]$ is equal to the parity of the total number of Dirac points in the whole BZ. Meanwhile, for the surface $S_{\mathrm{tot}}$, we have
\begin{align}
    \mathcal{Q}[S_{\mathrm{tot}}]=n_{\mathrm{Ch}}[k_z=0]\ \ (\mathrm{mod}\ 2), \label{eq:3-7}
    \end{align}
where $n_{\mathrm{Ch}}[k_z=0]$ denotes the Chern number on the plane $k_z=0$. It is because, in Eq.~(\ref{eq:3-1}), the terms $P[l]$ for the four paths $(k_x, k_z)=(\pm\pi, \pi)$ and $(k_y,k_z)=(\pm\pi,\pi)$ cancel each other and the integral of the Berry connection over $(S_{\mathrm{tot}})_+$ is equal to $n_{\mathrm{Ch}}[k_z=0]$ due to the periodicity of $H(\bm{k})$ over the BZ. Then, we find the failure of the NN-like theorem, because the $\tilde{\Theta}$ symmetry does not force $n_{\mathrm{Ch}}[k_z=0]$ to be zero. 

We can understand the reason for the failure of the NN-like theorem as follows. The $Z_2$ charge $\mathcal{Q}$ is defined for Dirac points on the plane $k_z=\pi$, but not on the plane $k_z=0$. Therefore, a Dirac point can annihilate singly when it is transformed into a Weyl dipole and the Weyl dipole moves to the plane $k_z=0$. Then, the number of Dirac points changes by unity and thus can be odd. We give an example of Dirac systems with a single Dirac point in App.~\ref{sec:D}. 

We finally note that, when the system has some additional symmetries which force $n_{\mathrm{Ch}}[k_z=0]$ to be zero, such as $\mathcal{T}$ symmetry and $\mathcal{PT}$ symmetry, the NN-like theorem holds.

\begin{figure}
\includegraphics[width=\columnwidth]{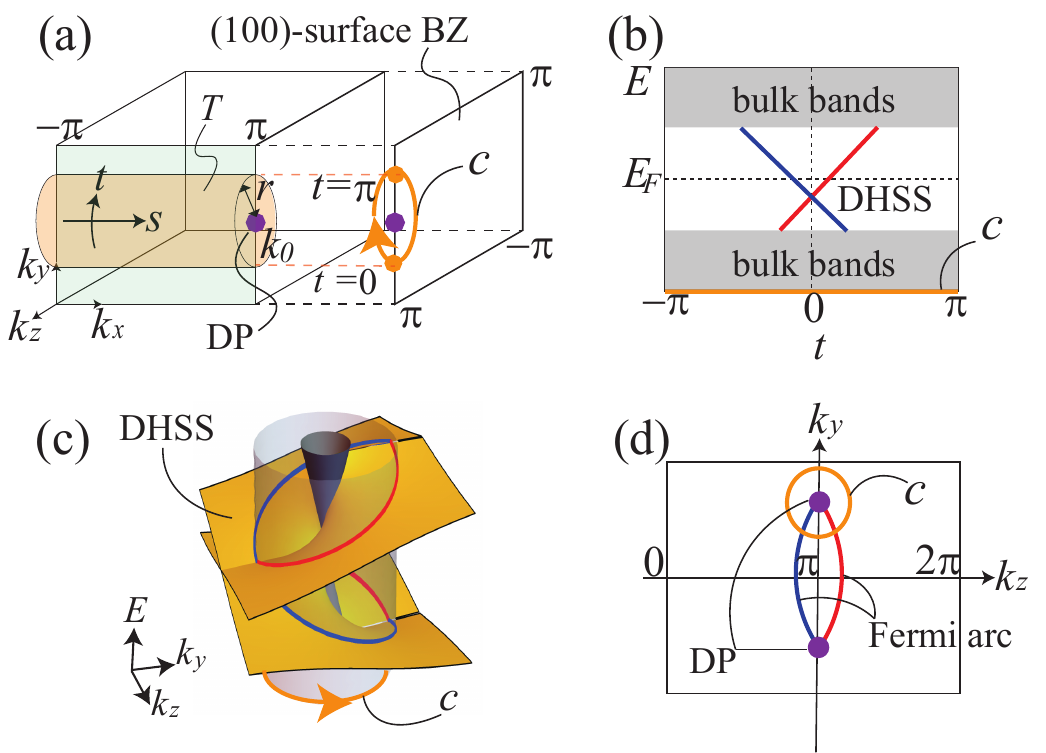}
\caption{\label{fig3}\small{Bulk-surface correspondence for $\mathcal{Q}[S]$. (a) Torus $T$ in the bulk BZ used for the bulk-surface correspondence. It is same as a torus in Fig.~\ref{fig2}(d) and it is projected onto a circle $c$ in the (100) surface BZ. (b) Surface band structure on $c$. When $T$ encloses a Dirac point, the dispersion of the surface states along $c$ must be helical, corresponding to the nontrivial value of $\mathcal{Q}[T]$. (c) Double-helicoid surface states. We can observe DHSSs by changing the radius $r$ of the torus $T$. (d) Double Fermi arcs on the (100) surface BZ. They connect the projections of Dirac points.}}
\end{figure}

\section{\label{sec:4}Bulk-surface correspondence}
In this section, we establish a bulk-surface correspondence, which claims that, in Dirac SMs with $\mathcal{GT}$ symmetry, a Dirac point in the bulk BZ leads to DHSSs on the surface BZ. This is already proposed in Ref.~\cite{Fang2016}, but without giving its proof. Therefore, we give a proof here. Briefly, we consider the $Z_2$ charge $\mathcal{Q}[S]$ with $S$ being the 2D torus $T$ defined in Sec.~\ref{sec:3} and show the bulk-surface correspondence for the 2D subsystem on this torus $T$. Details are as follows.

When we take the torus $T$, we have
\begin{align}
    \mathcal{Q}[T]= P_{\tilde{\Theta}}[t=\pi]-P_{\tilde{\Theta}}[t=0]\ \ (\mathrm{mod}\ 2), \label{eq:4-1}
\end{align}
as we have seen in Sec.~\ref{sec:3-2-1}. Meanwhile, we can regard the surface $T$ as a 2D BZ $[-\pi,\pi]\times[-\pi,\pi]$ in the $(s, t)$ parameter space. The antiunitary operator $\tilde{\Theta}$ acts on the 2D gapped system defined on this 2D BZ as $\tilde{\Theta} (s, t)=(-s, -t)$, and satisfies $\tilde{\Theta}^2=-1$ on the lines $t=0$ and $t=\pi$. Thus, similar to the bulk-surface correspondence for $\mathcal{T}$-protected $Z_2$ topological insulators \cite{FuKane2006, FuKane2007}, we deduce that the right hand side of Eq.~(\ref{eq:4-1}), the difference of the $\tilde{\Theta}$-polarization on two lines, distinguishes nontrivial surface states shown in Fig.~\ref{fig3}(b) from trivial surface states on $c$, where $c$ is the projection of $T$ onto the (100) surface BZ shown in Fig.~\ref{fig3}(a). Therefore, the nontrivial value of $\mathcal{Q}[T]$, which means there are an odd number of Dirac points inside of the torus $T$, leads to the surface states with helical dispersion along the circle $c$. By changing the value of the radius $r$, we conclude that a Dirac point leads to DHSSs, as shown in Fig.~\ref{fig3}(c). Moreover, we can observe the double Fermi arcs connecting the projection of Dirac points when there are two Dirac points in the BZ, as shown in Fig.~\ref{fig3}(d). 

The proof of the bulk-surface correspondence above is similar to that in Dirac SMs with time-reversal and reflection symmetries in Ref.~\cite{Morimoto2014}. In fact, both of them use $Z_2$ topology of 2D subsystems in the 3D bulk BZ to deduce the existence of nontrivial surface states; Ref.~\cite{Morimoto2014} uses the plane $k_y=\mathrm{const.}$ and we use the torus $T$. Meanwhile, as opposed to the case in Ref.~\cite{Morimoto2014}, we cannot use the plane $k_y=\mathrm{const.}$ to discuss the bulk-surface correspondence in our case. It is because $\tilde{\Theta}^2=-1$ does not hold on the plane $k_z=0$, and thus the plane $k_y=\mathrm{const.}$ cannot be characterized by the $Z_2$ invariant defined as the difference of the $\tilde{\Theta}$-polarization on two lines.

Finally, we note that the 2D gapped subsystem defined on the torus $T$ belongs to class A\(\rm{I}\hspace{-1pt}\rm{I}\) in the Altland-Zirnbauer symmetry classes~\cite{Alt1997} (see App.~\ref{sec:C}), and Eq.~(\ref{eq:4-1}) corresponds to the $Z_2$ invariant for 2D gapped systems with class A\(\rm{I}\hspace{-1pt}\rm{I}\)~\cite{FuKane2006, FuKane2007}. Therefore, we can also establish the bulk-surface correspondence for $\mathcal{Q}[T]$ through that for 2D gapped systems with class A\(\rm{I}\hspace{-1pt}\rm{I}\).

\section{\label{sec:5}Relationship with other topological invariants}
In this section, we compare the $Z_2$ charge $\mathcal{Q}[S]$ with other topological invariants under some additional symmetries. As we see below, the $Z_2$ charge $\mathcal{Q}[S]$ corresponds to the glide $Z_2$ invariant $\nu$ in spinless Dirac SMs with $\mathcal{G}$ and $\mathcal{T}$ symmetries, and the second SW number $w_2$ in spinless Dirac SMs with $\mathcal{GT}$ and $\mathcal{PT}$ symmetries.

\subsection{\label{sec:5-1}Glide $Z_2$ invariant}
In spinless Dirac SMs with $\mathcal{G}$ and $\mathcal{T}$ symmetries, which corresponds to MSG \#7.25 ($Pc1'$), the $Z_2$ charge $\mathcal{Q}[S]$ with a proper $S$ corresponds to the glide $Z_2$ invariant $\nu$, which characterizes the topological crystalline insulator phase protected by $\mathcal{G}$ symmetry~\cite{Shiozaki2015, Fang2015, Kim2019}. We obtain the result from the refinement of the comparison between $\tilde{\mathcal{Q}}$ and $\nu$ in Ref.~\cite{Zhang2022}.

\begin{figure}
\includegraphics[width=\columnwidth]{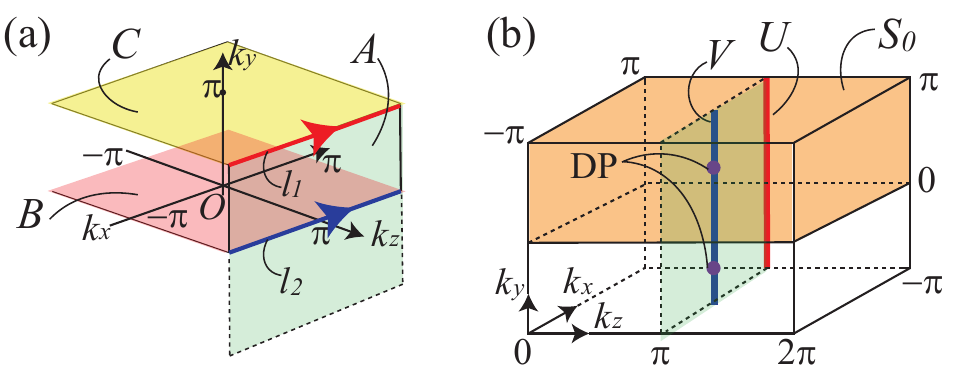}
\caption{\label{fig4}\small{Relationship between $\mathcal{Q}[S]$ and $\nu$. (a) Bulk BZ for MSG \#7.25 ($Pc1'$). The surfaces $\mathcal{A}=\{\bm{k} |-\pi\le k_x\le \pi,\ 0\le  k_y \le  \pi, \ k_z=\pi\}$, $\mathcal{B}=\{\bm{k} | -\pi\le k_x\le \pi,\ k_y=0,\ -\pi\le k_z\le \pi\}$, and $\mathcal{C}=\{\bm{k} | -\pi\le k_x\le \pi,\ k_y=\pi,\ -\pi\le k_z\le \pi\}$ are used to define the glide $Z_2$ invariant $\nu$. (b) Closed surface $S_0$. It is defined as the boundary of a half of the BZ: $S_0=\{\bm{k} | k_x=-\pi,\pi\ \mathrm{or}\ k_y=0,\pi\ \mathrm{or}\ k_z=0,2\pi\}$. Here, we shift the bulk BZ by $\pi$ along the $k_z$-direction.}}
\end{figure}

In such Dirac SMs, we can define $\mathcal{Q}[S]$. Meanwhile, if we add $\mathcal{T}$-breaking (but $\mathcal{G}$-preserving) perturbation, the Dirac SMs become gapped because such perturbation breaks the $\mathcal{GT}$ symmetry which protects Dirac points. In such a $\mathcal{G}$-protected gapped system, one can define the $Z_2$ invariant $\nu$ for the topological crystalline insulator phase protected by $\mathcal{G}$ symmetry as \cite{Shiozaki2015, Fang2015, Kim2019}
\begin{align} 
\nu=&\frac{1}{2\pi}\qty[\int_{\mathcal{A}}d\bm{S}\cdot\mathrm{rot}\bm{A}+\int_{\mathcal{B}-\mathcal{C}}d\bm{S}\cdot\mathrm{rot}\bm{A}^-] \notag \\
&-\frac{1}{\pi}\qty(\gamma^+[l_1]+\gamma^+[l_2]) \  (\mathrm{mod}\ 2), \label{eq:5-1}
\end{align}
where $\mathcal{A}, \mathcal{B}, \mathcal{C}$ are the regions in Fig.~\ref{fig4}(a), and $l_i\ (i=1,2)$ are lines introduced in Sec.~\ref{sec:2} with $k_1=\pi$ and $k_2=0$. $\bm{A}^{\pm}(\bm{k})$ and $\gamma^{\pm}[l]$ are defined similarly to Eqs.~(\ref{eq:2-4}),~(\ref{eq:2-5}), respectively, but the superscripts $i=\pm$ refer to the decomposition of the occupied states on the planes $k_z=0, \pi$ based on its eigenvalue of $\mathcal{G}$: $g_{\pm}(\bm{k})=\pm e^{-\frac{ik_z}{2}}$. 

Then, we have
\begin{align}
    \mathcal{Q}[S_0]=\nu \ \ (\mathrm{mod}\ 2), \label{eq:5-2}
\end{align}
when the Dirac SMs are gapped on the surface $S_0$, where $S_0$ is the boundary of a half of the BZ shown in Fig.~\ref{fig4}(b) (see App.~\ref{sec:B}). Here, Eq.~(\ref{eq:5-2}) means that the value of $\mathcal{Q}[S_0]$ for the original Dirac SMs is equal to the value of $\nu$ for the gapped system obtained by adding $\mathcal{T}$-breaking (but $\mathcal{G}$-preserving) perturbations. Therefore, we can obtain $\mathcal{G}$-protected topological crystalline insulators from spinless Dirac SMs with $\mathcal{G}$ and $\mathcal{T}$ symmetries.

\subsection{\label{sec:5-2}Second Stiefel-Whitney number}
In spinless Dirac SMs protected by $\mathcal{G}\mathcal{T}$ and $\mathcal{P}\mathcal{T}$ symmetries, which corresponds to MSGs \#13.68 ($P2/c'$) or \#14.78 ($P2_1/c'$), the $Z_2$ charge $\mathcal{Q}[S]$ is equal to the second SW number $w_2$ \cite{Fang2015-2, Ahn2018}. Interestingly, this correspondence leads to the coexistence of DHSSs and a hinge Fermi arc, as we see below.

Firstly, we briefly review the second SW number $w_2$. $w_2$ is a $Z_2$ invariant defined in 2D spinless gapped systems with $\mathcal{P}\mathcal{T}$. In 3D nodal line SMs, $Z_2$ nodal rings are characterized by the nontrivial value of $w_2$ for a sphere enclosing the nodal rings. We can calculate the value of $w_2$ on the sphere as follows~\cite{Fang2015-2}. Firstly, let us parameterize the sphere with the spherical coordinate $(\theta, \phi)\ (0\le\theta\le\pi, -\pi\le\phi\le\pi)$. Next, let us take real continuous gauges ($\mathcal{PT}\ket{u_n(\bm{k})}=\ket{u_n(\bm{k})}$) on the northern hemisphere $\ket{u^{N}_n(\theta, \phi)}\ (0\le\theta\le\frac{\pi}{2})$ and the southern hemisphere $\ket{u^{S}_n(\theta, \phi)}\ (\frac{\pi}{2}\le \theta \le \pi)$ respectively. Next, let us define $M(\phi)\in O(N_{\mathrm{occ}})$ as
$M_{mn}(\phi)=\braket{u^{N}_m(\frac{\pi}{2}, \phi)}{u^{S}_n(\frac{\pi}{2}, \phi)}$. 
Then, we have $w_2=N_{M(\phi)}\ (\mathrm{mod}\ 2)$, where $N_{M(\phi)}$ is the winding number of $M(\phi)$: $N_{M(\phi)}\in\pi_1(O(N_{\mathrm{occ}}))\cong Z_2\  (\mathrm{for}\ N_{\mathrm{occ}}>2),\ \cong Z\ (\mathrm{for}\ N_{\mathrm{occ}}=2)$.
We can also calculate the value of $w_2$ on a sphere or a torus by using the Wilson loop operator~\cite{Ahn2018}.

Then, we have
\begin{align}
\mathcal{Q}[S]=w_2[S]\ \ (\mathrm{mod}\ 2), \label{eq:5-4}
\end{align}
where $w_2[S]$ is the second SW number on the surface $S$ (see App.~\ref{sec:B}). In particular, the value of the second SW number on the plane $k_y=\mathrm{const}.$ changes by unity when the plane passes a Dirac point.
Therefore, in Dirac SMs with $\mathcal{GT}$ and $\mathcal{PT}$ symmetries, DHSSs and a hinge Fermi arc can coexist, because 2D gapped systems with $w_2=1$ are second-order topological insulators~\cite{Wang2020,Sheng2019, Lee2020, Chen2021}. We examine this in Sec.~\ref{sec:7-2}.

Finally, we note some points. First, the hinge Fermi arc is not topological in a strict sense. In fact, when the system does not possess chiral symmetry, the hinge states can spread over the bulk~\cite{Lee2020}. Second, the DHSSs is not a consequence of the nontrivial $w_2$ on the plane $k_y=\mathrm{const}.$, and just a consequence of the nontrivial $\mathcal{Q}[T]$. In fact, there is a weak SW insulator with $\mathcal{GT}$ and $\mathcal{PT}$ symmetries which does not have topological surface states, as we see in App.~\ref{sec:F}. Therefore, even if DHSSs appear on the projection of the plane $k_y=\mathrm{const}.$ with $w_2=1$ in some Dirac SMs, by considering the direct sum of the system and the weak SW insulator, we can change the value of $w_2$ to zero on the plane without affecting the construction of DHSSs. Third, although there are some previous studies which show that spinless Dirac SMs with $\mathcal{PT}$ symmetry can have topological surface states~\cite{Zhao2017, Wang2020}, these studies do not include our study. In fact, surface states in Dirac SMs with only $\mathcal{PT}$ symmetry are consequence of the nontrivial $w_2$~\cite{Zhao2017}, but DHSSs in the present study are not, as mentioned above. Moreover, the former can become gapped by adding some perturbations which change Dirac points to $Z_2$ nodal rings~\cite{Wang2020}. Meanwhile, the latter cannot be gapped by such perturbations as long as they preserve $\mathcal{GT}$ and $\mathcal{PT}$ symmetries, as we have seen in Sec.~\ref{sec:4}.

\section{\label{sec:6}Computation methods}
In this section, we develop computation methods for $\mathcal{Q}[S]$. While the definition of the $Z_2$ charge $\mathcal{Q}[S]$ given in Eq.~(\ref{eq:3-1}) is not convenient for the computation because of some integral terms, we can compute the value of $\mathcal{Q}[S]$ easily by using the Wilson loop methods or the Fu-Kane like formulas, as we see below.

\subsection{\label{sec:6-1}Wilson loop methods}
Similar to the $Z_2$ invariant for $\mathcal{T}$-protected insulators~\cite{Yu2011}, one can compute the value of $\mathcal{Q}[S]$ for $S=T,\ S^2$ by using the Wilson loop operator. 

\begin{figure}
\includegraphics[width=\columnwidth]{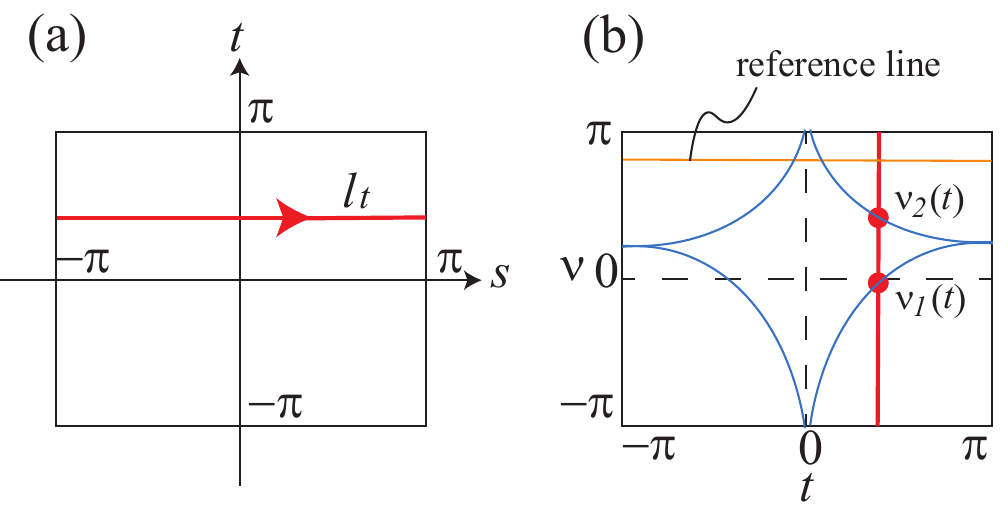}
\caption{\label{fig5}\small{Wilson loop spectrum. (a) Line $l_t$ in the torus $T$. We obtain $\nu_j(t)$ as the phase parts of the eigenvalues of the Wilson loop operator $\mathcal{W}[l_t]$ along $l_t$. (b) Example of the Wilson loop spectrum. Because bands intersect a reference line $\nu=\mathrm{const.}$ odd times within $0\le t\le\pi$, we have $\mathcal{Q}[T]=1$ for this spectrum. }}
\end{figure}

Firstly, let us review the definition of the Wilson loop operator. The Wilson loop operator $\mathcal{W}[l]$ for the curve $l$ in the BZ is defined as 
\begin{align}
  [\mathcal{W}[l]]_{mn}&=\mel{u_{m\bm{k}=l(\pi)}}{W[l]}{u_{n\bm{k}=l(-\pi)}},  \label{eq:6-0} \\
    W[l] 
    &=\lim_{N\to \infty}P_{l(\pi)}P_{l(\frac{N-1}{N}\pi)}\cdots P_{l(-\frac{N-1}{N}\pi)}P_{l(-\pi)}, \label{eq:6-1}
\end{align}
where $l$ is parameterized by $s\in[-\pi,\pi]$, and $P_{\bm{k}}$ is the projection operator to the space spanned by the occupied states: $P_{\bm{k}}=\sum_{n=1}^{N_{\mathrm{occ}}}\ketbra{u_{n\bm{k}}}{u_{n\bm{k}}}$. When $l$ is a closed curve, $\mathcal{W}[l]$ becomes a unitary matrix, and has eigenvalues $e^{i\nu_j}\ (-\pi<\nu_j\le\pi,\ j=1,2,\dots, N_{\mathrm{occ}})$. Now let us take a 2D closed surface $S$ in the BZ and parameterize the surface by $(s,t)\ (-\pi\le s\le\pi,\ -\pi\le t\le\pi)$, and consider the Wilson loop operator $\mathcal{W}[l_{t}]$, where $l_{t}$ is the line parallel to $s$-axis with fixed $t$, shown in Fig.~\ref{fig5}(a). Then, we obtain the spectrum of $\nu_j(t)$: the phase parts of the eigenvalues of $\mathcal{W}[l_{t}]$, as shown in Fig.~\ref{fig5}(b).

Next, we explain the computation method of $\mathcal{Q}[S]\ (S=T,\ \mathrm{or}\  S^2)$ using the Wilson loop operator. The torus $T$ and the sphere $S^2$ are parameterized by $(s,t)$ as described in Sec.~\ref{sec:3}. Therefore, we can obtain the Wilson loop spectrum for $T,\ S^2$ as explained above, and we have
\begin{align}
    \mathcal{Q}[S]=M_n\ \ \ (\mathrm{mod}\ 2) \label{eq:6-2}
\end{align}
for $S=T\ \mathrm{and}\ S^2$, where $M_n$ is the winding number of the Berry phase $\gamma[l_{t}]$ defined as
\begin{align}
2\pi M_n=\int_0^{\pi}\partial_{t}\gamma[l_{t}]dt-\sum_{i=1}^{N_{\mathrm{occ}}}\qty(\nu_i(\pi)-\nu_i(0)). \label{eq:6-3}
\end{align}
From Eq.~(\ref{eq:6-2}), we can see that the value of the $Z_2$ charge becomes nontrivial if and only if bands intersect a reference line $\nu=\mathrm{const.}$ odd times within $0\le t\le\pi$.

\subsection{\label{sec:6-2}Fu-Kane like formula}
In spinful Dirac SMs with $\mathcal{GT}$ and $\mathcal{PT}$ symmetries, which corresponds to MSGs \#13.68 ($P2/c'$), and \#14.78 ($P2_1/c'$), we obtain Fu-Kane like formulas for $\mathcal{Q}[S]$, i.e., $\mathcal{Q}[S]$ is expressed as the products of the eigenvalues of symmetry operators at high-symmetry points in $k$-space.

It is already discussed for $\mathcal{Q}$ in Ref.~\cite{Fang2016}, and for $\tilde{\mathcal{Q}}$ in Ref.~\cite{Zhang2022}, and we show the similar formula for $\mathcal{Q}[S]$, which is newly defined in this paper. Since the derivation is almost the same,  we only show the resulting formula for $\mathcal{Q}[S]$ here. For MSG \#13.68 ($P2/c'$), we have
\begin{align}
\mathcal{Q}[S]=\prod_{l\in\partial S_+}\prod_{i=1}^{N_{\mathrm{occ}}/2}\frac{\zeta_i[l(\pi)]}{\zeta_i[l(0)]},\label{eq:6-4} 
\end{align}
where $\zeta_i(\bm{k})$ is the $C_2=\{C_{2y}|0 0 \frac{1}{2}\}$ eigenvalue of the $i$-th occupied band at the $C_2$-invariant $\bm{k}$, and for MSG \#14.78 ($P2_1/c'$), we have
\begin{align}
\mathcal{Q}[S]=\prod_{l\in\partial S_+}\prod_{i=1}^{N_{\mathrm{occ}}/2}e^{i\frac{l(\pi)_y-l(0)_y}{2}}\frac{\xi_i[l(\pi)]}{\xi_i[l(0)]},
\label{eq:6-5}
\end{align}
where $\xi_i(\bm{k})$ is the $S_y=\{C_{2y}|0 \frac{1}{2} \frac{1}{2}\}$ eigenvalue of the $i$-th occupied band at the $S_y$-invariant $\bm{k}$. Equations (\ref{eq:6-4}) and (\ref{eq:6-5}) show that Dirac points appear as a consequence of the band inversion between bands with different $C_2$ or $S_y$ eigenvalues in spinful Dirac SMs with the MSGs.

We note that orthorhombic CuMnAs, a candidate of magnetic Dirac SMs proposed in Ref.~\cite{Tang2016}, belongs to this category. When the spin-orbital coupling is turned on and magnetic moments on Mn atoms are aligned along the $z$-axis, the symmetry of orthorhombic CuMnAs contains $\mathcal{PT}$ and $\mathcal{G}_z\mathcal{T}=\{M_z|\frac{1}{2} 0 \frac{1}{2}\}$. Then, $ab\ initio$ results in Ref.~\cite{Tang2016} show that there are Dirac points protected by the band inversion on the plane $k_x=\pi$, and there are gapless surface states on the (010) surface BZ. The results above indicate that Dirac points in the bulk BZ of orthorhombic CuMnAs are $Z_2$-charged, and thus it explains the reason for the emergence of DHSSs in orthorhombic CuMnAs from the bulk-surface correspondence discussed in Sec.~\ref{sec:4}.

\section{\label{sec:7}Example}

\begin{figure*}
\includegraphics[width=2\columnwidth]{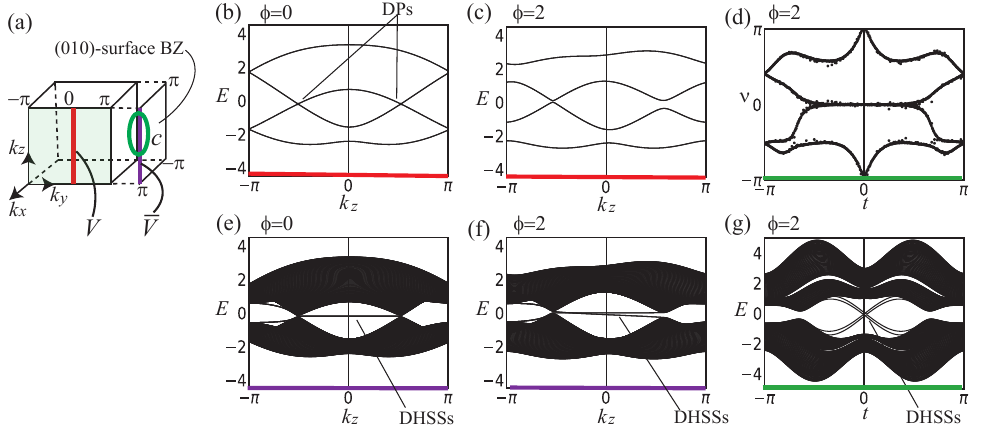}
\caption{\label{fig6}\small{Band structures of $H^{(A)}(\bm{k})$. Parameters are $t_1=1,\ t_2=0.8,\  t_3=0.5,\ t_4=0.3,\  t_5=0.8,\ t_6=0.3,\ t_7=0.7,\ t_8=0.3,\ m_x=0.3$, and $m_z=0.5$. (a) BZ for MSG \#7.26 ($Pc'$). (b) Bulk band structure on the line $V$ (red line) when $\phi=0$. Two Dirac points exist on the line. (c) Bulk band structure on $V$ when $\phi=2$. The Dirac points in (b) now split into pairs of two Weyl points. They exist out of the plane $k_z=\pi$, and thus the system is gapped on $V$. (d) Wilson loop spectrum for the torus $T$ defined in Sec.~\ref{sec:3} with $k_0=\pi/2$, $r=1$, and $\phi=2$. The spectrum shows $\mathcal{Q}[T]=1$. (e),(f) Band structures of the slab with (010) surface on the line $\bar{V}$ (purple line) when (e) $\phi=0$ and (f) $\phi=2$, respectively. DHSSs appear around the projection of Dirac points and pairs of Weyl points. (g) Band structure of the slab with (010) surface on the circle $c$ (green line) defined in Sec.~\ref{sec:4} with $k_0=\pi/2$, $r=1$, and $\phi=2$. Dispersion of DHSSs along the circle is helical. These nontrivial surface states are the consequence of the nontrivial value of $\mathcal{Q}[T]$.}}
\end{figure*}

\subsection{\label{sec:7-1}Spinful tight-binding model for MSG $\#7.26$}
We demonstrate the bulk-surface correspondence for $\mathcal{Q}[T]$ discussed in Sec.~\ref{sec:4} by constructing a spinful tight-binding model for MSG \#7.26 ($Pc'$) with eight bands. Firstly, we consider four sites within the unit cell as $A_1:(X,\frac{1}{4},Z),\ A_2:(X+\frac{1}{2},\frac{1}{4},-Z+\frac{1}{2}),\ A_3:(-X,\frac{3}{4},-Z),\ A_4:(-X+\frac{1}{2},\frac{3}{4},Z+\frac{1}{2})$ where $X$ and $Z$ are constants with $0<X<1,\ 0<Z< 1$, and put magnetic moments at each site as $M_1=(m_x, 0, m_z),\ M_2=(m_x, 0, -m_z),\ M_3=(-m_x, 0, -m_z),\ M_4=(-m_x, 0, m_z)$, respectively. Next, we introduce spin-independent real hopping $t_{1}$ between $A_1$ and $A_2$, $t_{2}$ between $A_3$ and $A_4$, $t_{3}$ between $A_4$ and $A_1$, and $t_{4}$ between $A_2$ and $A_3$. We also introduce spin-dependent real hopping $t_5$ between $A_1$ and $A_4$, and $t_6$ between $A_2$ and $A_3$. Furthermore, we introduce real hopping with a spin flip $t_7$ between $A_1$ and $A_4$, and $t_8$ between $A_2$ and $A_3$. Lastly, we introduce a complex phase $e^{i\phi}\ (\phi\in\mathbb{R})$ to $t_3$ and $t_4$. Then we obtain a tight-binding model $H^{(A)}(\bm{k})$, whose explicit form is given in App.~\ref{sec:D}. We assume half-filling, which means that four bands are occupied. When $\phi=0,\pi$, the tight-binding model is the same as a tight-binding model with MSG \#62.449 ($Pn'm'a'$) constructed in Ref.~\cite{Hara2023}. When $\phi\neq 0, \pi$, the tight-binding model only has the symmetry $\mathcal{G}_z\mathcal{T}=\{M_z | \frac{1}{2} 0 \frac{1}{2}\}'$ in addition to the lattice translation symmetry. 

Let us observe the band structures of $H^{(A)}(\bm{k})$. Firstly, we show the bulk band structure. When $\phi=0$, we can find two Dirac points on the line $V$ as shown in Fig.~\ref{fig6}(b). Meanwhile, when $\phi\neq 0$, these Dirac points split into pairs of two Weyl points as shown in Fig.~\ref{fig6}(c). Still in this case, we have $\mathcal{Q}[T]=1$ for a torus $T$ enclosing one of the pair of two Weyl points, as shown in Fig.~\ref{fig6}(d).

Next, we show the surface band structure. On the (010) surface, nontrivial surface states appear around the projection of the Dirac points when $\phi=0$ as shown in Fig.~\ref{fig6}(e). Even when  $\phi\neq 0$, these nontrivial surface states still appear around the projection of the pairs of Weyl points as shown in Fig.~\ref{fig6}(f). Moreover, when we take a circle enclosing the projection of one of the pairs of Weyl points, we can see that the dispersion of the surface states along the circle is helical, as shown in Fig.~\ref{fig6}(g). Therefore, we conclude that the observed surface states in Figs.~\ref{fig6}(e-g) are DHSSs. They are the consequence of the nontrivial value of $\mathcal{Q}[T]$ as discussed in Sec.~\ref{sec:4}.

\begin{figure*}
\includegraphics[width=2\columnwidth]{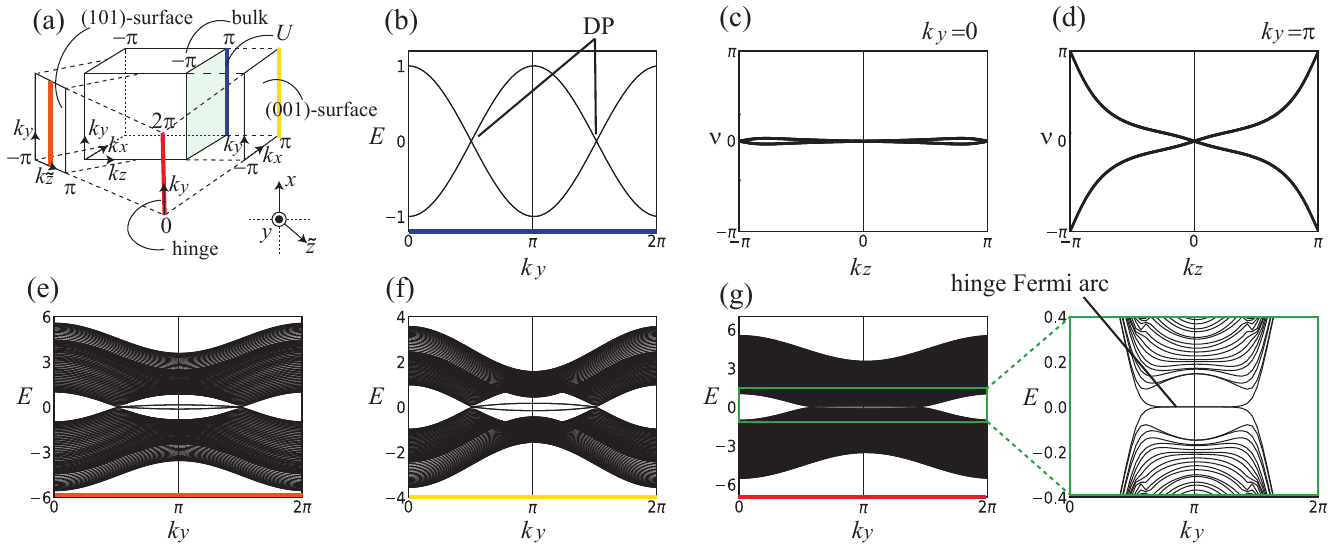}
\caption{\label{fig7}\small{Band structures of the model $H^{(B)}(\bm{k})$. Parameters are $m=2$, and $t=0.2$. (a) BZ for MSG \#13.68 ($P2/c'$) (b) Bulk band structure on the line $(\pi, k_y, \pi)\ (0<k_y<2\pi)$. We can find two Dirac points on the line. (c),(d) Wilson loop spectrum for the plane $k_y=0$ and $k_y=\pi$, respectively. We can find $w_2=0$ when $k_y=0$ and $w_2=1$ when $k_y=\pi$. (e),(f) Band structures of the slab with (010) surface on the line $k_{\tilde{z}}=0$, and with (001) surface on the line $k_{x}=\pi$, respectively. We can observe surface states around $E=0$ but these states are not topological because $\tilde{\Theta}$ symmetry is not protected on the surfaces. (g) Band structure of the hinge, which is periodic along the [010] direction, and finite-size along the [101] and [001] directions. The hinge Fermi arc appears and connects the projections of the Dirac points, corresponding to the nontrivial value of $w_2$ on the plane $k_y=\pi$.}}
\end{figure*}

\subsection{\label{sec:7-2}Spinless tight-binding model for MSG $\# 13.68$}
Next, we demonstrate the coexistence of the DHSSs and the hinge Fermi arc discussed in Sec.~\ref{sec:5-2}, by constructing a spinless tight-binding model for MSG $\# 13.68$. We put four sites $A_1: (X,Y,0)$, $A_2: (-X,Y,-\frac{1}{2})$, $A_3: (X, -Y, \frac{1}{2})$, and $A_4: (-X, -Y, 0)\  (0<X, Y<1)$, and take the basis set $[\psi_{A_1}, \psi_{A_2}, \psi_{A_3}, \psi_{A_4}]=[\psi_{A_1}, \mathcal{PT}\tilde{\Theta}\psi_{A_1}, \tilde{\Theta}\psi_{A_1}, \mathcal{PT}\psi_{A_1}]$. By introducing real hopping between the sites, we have a 3D tight-binding model whose Bloch Hamiltonian $H^{(B)}(\bm{k})$ is given by
\begin{align}
    &H^{(B)}(\bm{k}) \notag \\
    &=(m-\cos k_x-\frac{1}{2}\cos k_y -\cos k_z)\sigma_0\tau_1+\sin k_x\sigma_0\tau_2 \notag \\
    &\ +\sin k_z\sigma_3\tau_3+t(1+\cos k_z)\sigma_1\tau_0+t\sin k_z\sigma_2\tau_0 \notag \\
    &\ \ -t\sin k_z\sigma_1\tau_1+t(1+\cos k_z)\sigma_2\tau_1 \label{eq:7-1}
\end{align}
under the basis set, where $m$ and $t$ are real parameters, $\sigma_0$ and $\tau_0$ are the $2\times 2$ unit matrices, and $\sigma_j$ and $\tau_j$ are the Pauli matrices referring to the four sublattice sites, i.e. $(\sigma_3, \tau_3)=(+,+),\ (+,-),\ (-,+),\ (-,-)$ for the $A_1,\ A_2,\ A_3$, and $A_4$ sublattices, respectively. 
The tight-binding model has $\tilde{\Theta}$ and $\mathcal{PT}$ symmetries represented by
\begin{align}
    \tilde{\Theta} (\bm{k})=\mqty(0 & e^{ik_z} \\ 1  & 0 )\tau_0K,  \ \ &\ \mathcal{PT}(\bm{k})=\sigma_1\tau_1K,  \label{eq:7-2}
\end{align}
respectively, where $K$ is the complex conjugate operator. 

We observe the band structures of $H^{(B)}(\bm{k})$. Firstly, we show the bulk band structure. The tight-binding model has two Dirac points on the line $U$ as shown in Fig.~\ref{fig7}(b). Moreover, $w_2=0$ for the plane $k_y=0$ as shown in Fig.~\ref{fig7}(c), and $w_2=1$ for the plane $k_y=\pi$ as shown in Fig.~\ref{fig7}(d). Thus, these two Dirac points are $Z_2$-charged: $\mathcal{Q}=w_2=1\ (\mathrm{mod}\ 2)$. 

Next, we show the surface and the hinge band structure. On the (100) surface, DHSSs connecting the projection of two Dirac points appear due to the nontrivial value of $\mathcal{Q}$. Meanwhile, on the (101) surface or the (001) surface, although surface states are observed, one can find that these surface states are not gapless, and thus DHSSs do not appear on the surfaces, as shown in Fig.~\ref{fig7}(e) and Fig.~\ref{fig7}(f), respectively. It is because the $\tilde{\Theta}=\{M_y|00\frac{1}{2}\}'$ symmetry is not protected on these surfaces. Meanwhile, by calculating band structure in a cylinder geometry, which is periodic along the [010] direction, and finite-size along the [101] and [001] directions, the hinge Fermi arc appears as shown in Fig.~\ref{fig7}(g), as a consequence of the nontrivial value of the second SW number on the plane $k_y=\pi$. These observations are consistent with our discussions in Sec.~\ref{sec:5-2}.

\section{\label{sec:8}Conclusion}
In this paper, we refined the theory of $\mathcal{GT}$-protected Dirac semimetals. We firstly pointed out that the $Z_2$ charge $\tilde{\mathcal{Q}}$ defined in Ref.~\cite{Zhang2022} is ill-defined, and defined another $Z_2$ charge $\mathcal{Q}[S]$ for a surface $S$ satisfying the given conditions. When we take $S$ to be the torus $T$, $\mathcal{Q}[T]$ is almost the same as $\tilde{\mathcal{Q}}$, but with the corrected gauge conditions. On the other hand, when we take $S$ to be the sphere $S^2$, $\mathcal{Q}[S^2]$ is equal to the $Z_2$ charge $\mathcal{Q}$ defined in Ref.~\cite{Fang2016}. Therefore, we can unify the discussions on the $Z_2$ charges in Refs.~\cite{Fang2016, Zhang2022}. Next, by using the newly defined $Z_2$ charge, we established the bulk-surface correspondence, which claims that a Dirac point in the bulk Brillouin zone leads to double-helicoid surface states around its projections on the surface Brillouin zone. Furthermore, we showed that the $Z_2$ charge $\mathcal{Q}[S]$ is equal to the glide $Z_2$ invariant for $\mathcal{G}$-protected topological crystalline insulators in spinless Dirac semimetals with $\mathcal{G}$ and $\mathcal{T}$ symmetries, and is equal to the second Stiefel-Whitney number for $\mathcal{PT}$-protected nodal line semimetals in spinless Dirac semimetals with $\mathcal{GT}$ and $\mathcal{PT}$ symmetries.  We also discussed computation methods for $\mathcal{Q}[S]$ using Wilson loop methods and Fu-Kane like formulas.

For future works, it is desired to uncover unique physical properties which appear in $\mathcal{GT}$-protected Dirac semimetals but do not appear in conventional Dirac semimetals such as Na$_2$Bi and Cd$_2$As$_3$, due to the robustness of double-helicoid surface states under symmetry preserving perturbations. Because orthorhombic CuMnAs is expected to have such DHSSs, as pointed out in Ref.~\cite{Tang2016}, this can be achieved by extensive calculations or experiments on orthorhombic CuMnAs or the family of the material.

\begin{acknowledgments}
We acknowledge the support by Japan Society for the Promotion of Science (JSPS), KAKENHI Grant No.JP22K18687, No.JP22H00108, and No.JP24H02231.  We also acknowledge helpful comments from T. Zhang.
\end{acknowledgments}

\appendix

\section{\label{sec:A}Gauge-dependency of $\tilde{\mathcal{Q}}$}
In this section, we give a gauge transformation which alters the value of $\tilde{\mathcal{Q}}$ by unity, as mentioned in Sec.~\ref{sec:3}. Below, we assume a Dirac point exists at $\bm{k}=(0,0,\pi)$, and $0 <k_1 <\pi$,\ $-\pi< k_2 <0$. We consider $U(1)$ gauge transformation over one of the occupied bands.

\begin{figure}
\includegraphics[width=\columnwidth, pagebox=cropbox, clip]{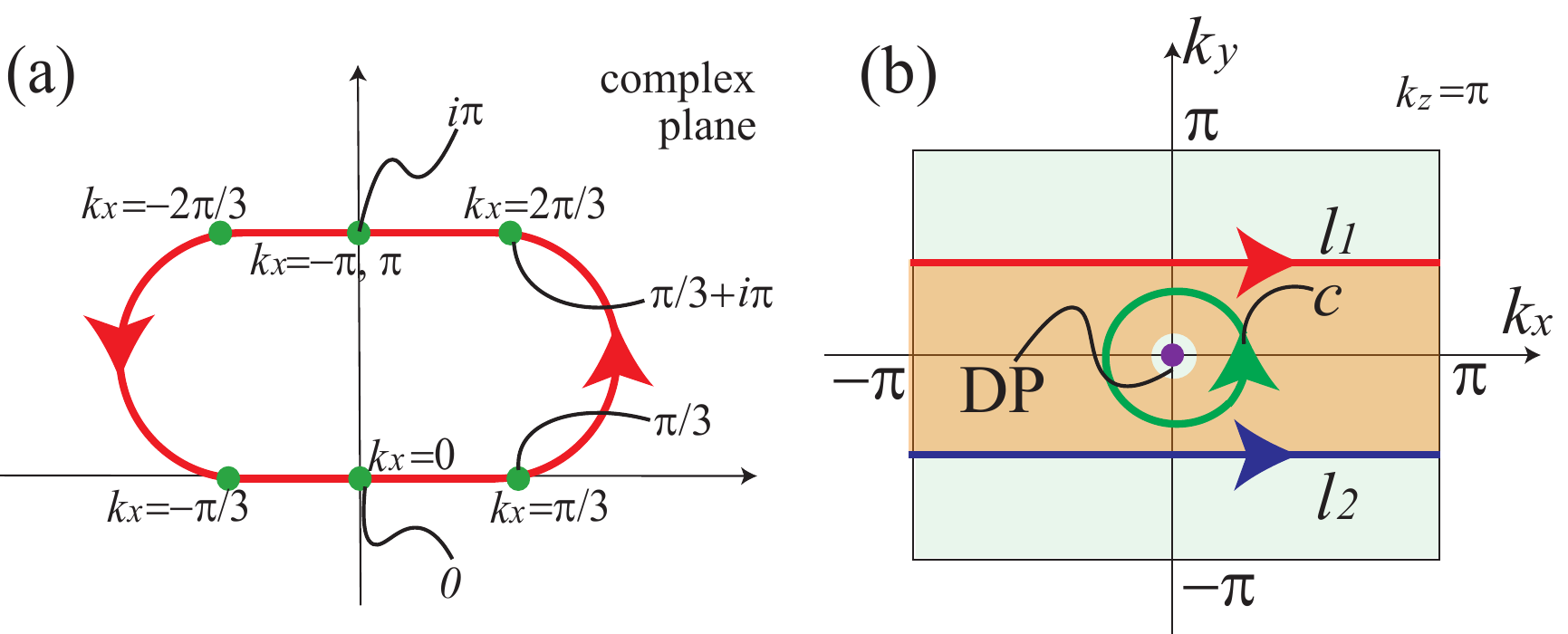}
\caption{\label{figA}\small{Gauge transformation changing the value of $\tilde{\mathcal{Q}}$ by unity. (a) Path of $z(k_x, 0)\ (-\pi\le k_x\le\pi)$. (b) Lines $c$, $l_1$, and $l_2$. $c$ is transformable into $-l_1+l_2$ without closing the gap.}}
\end{figure}

We define a gauge transformation $U(\bm{k})\ (\ket{u_{\bm{k}}}\mapsto U(\bm{k})\ket{u_{\bm{k}}})$  over the plane $\{\bm{k}| -\pi\le k_x\le \pi,\ k_2\le k_y\le k_1,\ k_z=\pi\}$ as
$U(k_x,k_y,k_z=\pi)=\frac{z(k_x,k_y)}{|z(k_x,k_y)|}$ where
\begin{widetext}
\begin{equation}
  z(k_x,k_y)=
  \begin{cases}
    (-\pi-k_x)+i(\pi+k_y) & \text{if $-\pi\le k_x\le-\frac{2\pi}{3}$,} \\
    (-\frac{\pi}{3}-\frac{\pi}{2}\sin 3k_x)+i(\frac{\pi}{2}+\frac{\pi}{2}\cos 3k_x+k_y)                & \text{if $-\frac{2\pi}{3}\le k_x \le -\frac{\pi}{3}$,} \\
    k_x+ik_y       & \text{if $-\frac{\pi}{3}\le k_x\le\frac{\pi}{3}$,} \\
    (\frac{\pi}{3}-\frac{\pi}{2}\sin 3k_x)+i(\frac{\pi}{2}+\frac{\pi}{2}\cos 3k_x+k_y)                & \text{if $\frac{\pi}{3}\le k_x \le\frac{2\pi}{3}$,} \\
     (\pi-k_x)+i(\pi+k_y) & \text{if $\frac{2\pi}{3}\le k_x\le\pi$,}
  \end{cases}
\end{equation}
\end{widetext}
(see Fig.~\ref{figA}(a)). 
Because $z(k_x,k_y)=0$ iff $(k_x,k_y)=(0,0)$, and
$z(k_x, k_y)$ satisfies the periodicity of the BZ along the $k_x$ direction, $U(\bm{k})$ is a smooth gauge transformation 
over the plane except for $\bm{k}=(0,0,\pi)$. 

We prove that $U(\bm{k})$ changes the value of $\tilde{\mathcal{Q}}$ by unity. We define the circle $c$ as $c(t)=(r\cos t, r\sin t, \pi)\ (-\pi\le t\le\pi)$, where the radius $r$ is taken to be less than $\frac{\pi}{3}$. Then, $U(\bm{k})$ satisfies $U(k_x, k_y, k_z=\pi)=e^{iArg(k_x+ik_y)}$ on $c$. Thus, $U(\bm{k})$ change the value of $\frac{1}{2\pi}\gamma[c]$ by unity.
Then, $U(\bm{k})$ also changes the value of $\frac{1}{2\pi}\gamma[-l_1+l_2]$ by unity, because $c$ is transformable into $-l_1+l_2$ without closing the gap, where $l_i\ (i=1,2)$ are shown in Fig.~\ref{figA}(b). Meanwhile, $U(\bm{k})$ does not change the value of $2P[l_i]\ (i=1,2)$ modulo 2. Therefore, $U(\bm{k})$ change the value of
\begin{align}
    \tilde{\mathcal{Q}}=\frac{1}{2\pi}\qty(\gamma[l_1]-\gamma[l_2])-2(P[l_1]-P[l_2])\ \  (\mathrm{mod}\ 2)
\end{align}
by unity. This result means that $\tilde{\mathcal{Q}}$ is not gauge-independent and is not a well-defined topological charge.

\section{\label{sec:B}Proofs of some properties of $\mathcal{Q}[S]$}
In this section, we give proofs of some properties of $\mathcal{Q}[S]$ omitted in Secs.~\ref{sec:3}, \ref{sec:5}, and \ref{sec:6}. Below, we assume a curve $l$ is closed under $\tilde{\Theta}$, exists on the plane $k_z=\pi$, and parameterized by $s\in[-\pi,\pi]$ so that $l(-s)=\tilde{\Theta}l(s)$. Then the curve $l_+$ is a half of $l$ with $0\le s\le \pi$ (see Fig.~\ref{fig2}(e)).

\paragraph{Gauge-independency of $P[l]$} 
\ \ Let us consider a $U(N_{\mathrm{occ}})$ gauge transformation $U(\bm{k})\ (\ket{u_{m\bm{k}}}\mapsto \ket{u_{n\bm{k}}}'=U(\bm{k})_{mn}\ket{u_{m\bm{k}}}\ (m,n=1,2,\dots, N_{\mathrm{occ}}))$. Under the gauge transformation, the Berry connection $\bm{A}(\bm{k})$ changes into $\bm{A}(\bm{k})+i\bm{\nabla}_{\bm{k}}\log\mathrm{det}U(\bm{k})$. Meanwhile, the sewing matrix $\omega(\bm{k})$ changes into $U(\tilde{\Theta}\bm{k})^{\dagger}\omega (\bm{k})U(\bm{k})^{*}$, and its Pfaffian $\mathrm{Pf}\omega(\bm{k})$ changes into $\mathrm{det}U(\bm{k})^{\dagger}\mathrm{Pf}\omega (\bm{k})$ for $\tilde{\Theta}$-invariant $\bm{k}$. By using these relations, we can see that $P[l]$ is gauge independent modulo 1.

\paragraph{Proof of Eq.(\ref{eq:3-4})}
Firstly, due to the $\tilde{\Theta}$ symmetry, we have
\begin{align}
    \bm{A}(\tilde{\Theta}\bm{k})=-\tilde{\Theta}\cdot\qty[\bm{A}(\bm{k})+i\bm{\nabla}_{\bm{k}}\log\mathrm{det}\omega (\bm{k})] \label{eq:B-1}
\end{align}
on the plane $k_z=\pi$. By using this equation, we have
\begin{align}
\int_{-\tilde{\Theta}l_+}d\bm{k}\cdot\bm{A}
&=\int_{l_+}d\bm{k}\cdot\bm{A}+i\log\qty(\frac{\mathrm{det}\omega [l(\pi)]}{\mathrm{det}\omega [l(0)]}). \label{eq:B-2}
\end{align}
By using Eq.~(\ref{eq:B-2}), we have
\begin{align}
&\frac{1}{2\pi}\gamma[l]-2P[l] \notag \\
&=\frac{i}{\pi}\log\qty(\frac{\sqrt{\mathrm{det}\omega [l(\pi)]}}{\mathrm{Pf}\omega [l(\pi)]}\frac{\mathrm{Pf}\omega [l(0)]}{\sqrt{\mathrm{det}\omega [l(0)]}}) 
\end{align}
Meanwhile, by taking a smooth gauge over $S$, we have
\begin{align}
    \mathcal{Q}[S]=\sum_{l\in\partial S_h}\qty(\frac{1}{2\pi}\gamma[l]-2P[l])\ \ (\mathrm{mod}\ 2).
\end{align}
Therefore, we finally have Eq.~(\ref{eq:3-4}).

\paragraph{Proof of Eq.(\ref{eq:3-3})}
Firstly, due to the $\tilde{\Theta}$ symmetry, we have
\begin{align}
\int_{-\tilde{\Theta}l_+}d\bm{k}\cdot\bm{A}^{(\alpha)} =\int_{l_+}d\bm{k}\cdot\bm{A}^{(\beta)}+i\log\qty(\frac{\mathrm{det}\omega^{(\beta)} [l(\pi)]}{\mathrm{det}\omega^{(\beta)} [l(0)]}),
\label{eq:B-3}
\end{align}
where $\omega^{(i)}(\bm{k})\ (i=\alpha, \beta)$ are $N_{\mathrm{occ}}\times N_{\mathrm{occ}}$ matrices defined as
$[\omega^{(\alpha)}(\bm{k})]_{mn}=\mel{u_{m\tilde{\Theta}\bm{k}}^{(\beta)}}{\tilde{\Theta}}{u_{n\bm{k}}^{(\alpha)}}$, and
$[\omega^{(\beta)}(\bm{k})]_{mn}=\mel{u_{m\tilde{\Theta}\bm{k}}^{(\alpha)}}{\tilde{\Theta}}{u_{n\bm{k}}^{(\beta)}}$, respectively.
When $\bm{k}$ is $\tilde{\Theta}$-invariant, $\omega^{(\alpha)}(\bm{k})=-[\omega^{(\beta)}(\bm{k})]^T$. Thus, we have
\begin{align}
\omega(\bm{k})=\mqty(O & \omega^{(\alpha)}(\bm{k}) \\ \omega^{(\beta)}(\bm{k}) & O) =\mqty(O & -[\omega^{(\beta)}(\bm{k})]^T \\ \omega^{(\beta)}(\bm{k}) & O).  \label{eq:B-4}
\end{align}
 Meanwhile, for a $n\times n$ square matrix $B$, 
\begin{align}
    \mathrm{Pf}\mqty( O & -B^T \\ B & O )=(-1)^{\frac{n(n-1)}{2}}\mathrm{det}B \label{eq:B-5}
\end{align}
holds. By using Eqs.~(\ref{eq:B-3})-(\ref{eq:B-5}), we finally have
\begin{align}
    &\gamma^{(\alpha)}[l] \notag \\   &=\int_{l_+}d\bm{k}\cdot\qty[\bm{A}^{(\alpha)}+\bm{A}^{(\beta)}]+i\log\qty(\frac{\mathrm{det}\omega^{(\beta)}[l(\pi)]}{\mathrm{det}\omega^{(\beta)}[l(0)]}) \notag \\
    &=2\pi P[l]\ \ \ (\mathrm{mod}\ 2\pi).
\end{align}

\paragraph{Proof of Eq.~(\ref{eq:5-2})}
We give a proof of Eq.~(\ref{eq:5-2}) omitted in Sec.~\ref{sec:5-1}. Firstly, we have $\mathcal{Q}[S_0]=\nu[(S_0)_+]\ (\mathrm{mod}\ 2)$, where $S_0=\{\bm{k} | k_x=-\pi,\pi\ \mathrm{or}\ k_y=0,\pi\ \mathrm{or}\ k_z=0,2\pi\}$ and $\nu[(S_0)_+]$ is defined by changing the integral region of the first term in Eq.~(\ref{eq:5-1}) from the plane $\mathcal{A}$ to the surface $(S_0)_+$ (a half of $S_0$ with $k_z\ge\pi$). This is because we have $\int_{\mathcal{B-C}}d\bm{S}\cdot\mathrm{rot}\bm{A}^-=0$ due to the $\mathcal{T}$ symmetry, and we have $\gamma^+[l_i]=2\pi P[l_i]\ (i=1,2)$ from Eq.~(\ref{eq:3-3}). Next, when we add the $\mathcal{T}$-breaking perturbation so that gapless nodes do not pass through the surface $S_0$, the value of $\nu[(S_0)_+]$ does not change under the perturbation, because the value of $\nu[(S_0)_+]$ is an integer. Moreover, we have $\nu=\nu[(S_0)_+]$ when the system is gapped, because the surface $(S_0)_+$ is transformable into $\mathcal{A}$ without passing through gapless nodes. Combining these facts, we finally have Eq.~(\ref{eq:5-2}).

\paragraph{Proof of Eq.~(\ref{eq:5-4})}
We give a proof of Eq.~(\ref{eq:5-4}) omitted in Sec.~\ref{sec:5-2}. Because $S$ can be transformed into a sum of some spheres $S^2$ enclosing gapless nodes without passing through any gapless nodes, we only prove Eq.~(\ref{eq:5-4}) for $S=S^2$. To this end, we consider a spinless Dirac SMs with MSG \#13.68 ($P2/c'$) or MSG \#14.78 ($P2_1/c'$), and define $G=e^{-ik_z/2}\tilde{\Theta_y}$ for \#13.68 and $G=e^{i(k_y-k_z)/2}\tilde{\Theta}_y$ for \#14.48. Then, we have $G\mathcal{PT}=\mathcal{PT}G$. This means that when $\ket{u_n(\bm{k})}$ is real, $G\ket{u_n(G\bm{k})}$ is also real. Using this fact, we choose a smooth real gauge satisfying $\ket{u^{S}_n(\theta, \phi)}=G\ket{u^{N}_n(\pi-\theta, -\phi)}$. Then we have $M(\phi)=\omega_G(-\phi)$, where $[\omega_G(\phi)]_{mn}=\mel{u_m(\frac{\pi}{2}, -\phi)}{G}{u_n(\frac{\pi}{2},\phi)}$. Moreover, we have
\begin{align}
    (-1)^{N_{\omega_G(\phi)}}=\mathrm{Pf}[\omega_G(\phi=0)]\mathrm{Pf}[\omega_G(\phi=\pi)]. \label{eq:B-6}
\end{align}
(Proof is given below.) Meanwhile, because $\omega_G(\phi)$ is a real matrix, we have
\begin{align}
    (-1)^{\mathcal{Q}}=\mathrm{Pf}[\omega_G(\phi=0)]\mathrm{Pf}[\omega_G(\phi=\pi)] \label{eq:B-7}
\end{align}
from Eq.~(\ref{eq:2-3}). (We can calculate $\mathcal{Q}$ from $\omega_G(\bm{k})$ instead of $\omega(\bm{k})$.) Finally, combining Eqs.~(\ref{eq:B-6}) and (\ref{eq:B-7}), we have Eq.~(\ref{eq:5-4}).

We give a proof of Eq.~(\ref{eq:B-6}). Firstly, because $\omega_G(\phi)\in SO(N_{\mathrm{occ}})$, $\omega_G(\phi)$ can be continuously diagonalized into blocks of $SO(2)$ matrices $\omega_G^{(i)}(\phi)\ (i=1,2,\dots, N_{\mathrm{occ}}/2)$ for $\phi\in[0,\pi]$ by using $O(\phi)\in SO(N_{\mathrm{occ}})$:
\begin{align}
    \omega_G(\phi)=O(\phi)\mathrm{diag}[\omega_G^{(1)}(\phi), \dots, \omega_G^{(N_{\mathrm{occ}}/2)}(\phi)]O(\phi)^T. \label{eq:B-8}
\end{align}
By using $\omega_G(-\phi)=-\omega_G(\phi)^T$, $\omega_G(\phi)$ can be diagonalized also for $\phi\in[-\pi,0]$ as Eq.~(\ref{eq:B-8}) with $O(-\phi)=O(\phi)\ (\phi\in[0,\pi])$. Then, we have $\omega_G^{(i)}(-\phi)=-\omega_G^{(i)}(\phi)^T$, and $N_{\omega_G}=N_{O^T\omega_GO}=\sum_iN_{\omega^{(i)}_G}\ (\mathrm{mod}\ 2)$. Therefore, we only have to show Eq.~(\ref{eq:B-6}) for $N_{\mathrm{occ}}=2$.

We consider the case where $N_{\mathrm{occ}}=2$. In the case, $\omega_G(0)=i\sigma_2$ or $-i\sigma_2$ because $\omega_G(0)=-\omega_G(0)^T$ and $\omega_G(0)\in SO(2)$. Below, We assume $\omega_G(0)=i\sigma_2$ because the case where $\omega_G(0)=-i\sigma_2$ can be reduced to this case by redefining $\omega_G'=-\omega_G$. When $N_{\mathrm{occ}}=2$, there is an isomorphism from $\mathbb{R}/2\pi\mathbb{Z}\cong S^1$ to $SO(2)$ given by 
\begin{align}
\theta\in S^1\mapsto \mqty(\sin\theta & \cos\theta \\ -\cos\theta & \sin\theta )\in SO(2).
\end{align}
Then, $\theta(0)=0$ and the condition $\omega_G(-\phi)=-\omega_G(\phi)^T$ reduces to $\theta(-\phi)=-\theta(\phi)$ ($\omega_G(\phi)\in SO(2)$ corresponds to $\theta(\phi)\in S^1$ under the isomorphism). Then, when we consider a lift of $\theta: [-\pi,\pi]\rightarrow S^1$ through the projection $p: \mathbb{R}\rightarrow S^1$ ($\tilde{\theta}: [-\pi,\pi]\rightarrow S^1$ s.t. $p\circ\tilde{\theta}=\theta$), we have $\tilde{\theta}(-\phi)=-\tilde{\theta}(\phi)$ (we set $\tilde{\theta}(0)=0$). Then we have $N_{\theta(\phi)}=\tilde{\theta}(\pi)-\tilde{\theta}(-\pi)=2\tilde{\theta}(\pi)$. Therefore, $N_{\omega_G(\phi)}=N_{\theta(\phi)}=1\ (\mathrm{mod}\ 2)$ iff $\tilde{\theta}(\pi)=(2n+1)\pi\ (n\in\mathbb{Z})$ iff $\omega_G(\pi)=-i\sigma_2$ iff $\mathrm{Pf}[\omega_G(\pi)]=-1$. Thus we finally have Eq.~(\ref{eq:B-6}) for $N_{\mathrm{occ}}=2$

\section{\label{sec:C}Another proof of the bulk-surface correspondence}
In this section, we prove that the 2D subsystems defined on the torus $T$ belong to class A\(\rm{I}\hspace{-1pt}\rm{I}\). This fact gives another proof of the bulk-surface correspondence as mentioned in Sec.~\ref{sec:4}. First, under a proper choice of the basis set, we have
\begin{align}
    \tilde{\Theta}(\bm{k}) &=\mqty(0 & e^{ik_z} \\ 1 & 0)\otimes I_{N}K \label{eq:C-1}
\end{align}
where $I_N$ is the $N\times N$ identity matrix, and $N$ is half of the number of the basis set. Next, let us define the operation $\tilde{\Theta}'(\bm{k})$ as
\begin{align}
    \tilde{\Theta}'(\bm{k})=e^{-i(k_z-\pi)/2}\tilde{\Theta}(\bm{k}). \label{eq:C-2}
\end{align}
Then, under the unitary transformation $U(\bm{k})$ $(H(\bm{k})\mapsto U(\bm{k})H(\bm{k})U(\bm{k})^{\dagger}$, $\tilde{\Theta}'(\bm{k})\mapsto U(\bm{k}')\tilde{\Theta}'(\bm{k})U(\bm{k})^{\dagger}$ where $\bm{k}'=(-k_x, k_y, 2\pi-k_z)\ )$ defined as
\begin{align}
    U(\bm{k})=\mqty(e^{i(k_z-\pi)/4} & 0 \\ 0 & e^{-i(k_z-\pi)/4})\otimes I_N,
\end{align}
we have
\begin{align}
\tilde{\Theta}'(\bm{k})H(\bm{k})\tilde{\Theta}'(\bm{k})^{-1}&=H(\bm{k}'), \\
    \tilde{\Theta}'(\bm{k}')\tilde{\Theta}'(\bm{k})&=-1, \\
      \tilde{\Theta}'(\bm{k})&=\mqty(0 & -1 \\ 1 & 0)\otimes I_NK.
\end{align}
Therefore, when we consider the 2D subsystem defined on the torus $T$, the system belongs to class A\(\rm{I}\hspace{-1pt}\rm{I}\). Here we note that $\tilde{\Theta}'(\bm{k})$ and $U(\bm{k})$ are uniquely defined on the torus $T$.

\section{\label{sec:D}Tight binding model with a single Dirac point}
In this section, we give a tight-binding model with a single Dirac point, as mentioned in Sec.~\ref{sec:3-2-3}. To this end, we construct a tight-binding model for MSG \#7.26 ($Pc'$) with four bands as follows. We put four sites in the unit cell; $A_1:(X,Y,0)$, $A_2:(X', Y', 0)$, $A_3:(X, -Y, 1/2)$, and $A_4:(X', -Y', 1/2)\ \ (0<X,X'<1,\ 0<Y,Y'<1/2)$, and take the basis set $[\psi_{A_1}, \psi_{A_2}, \psi_{A_3}, \psi_{A_4}]=[\psi_{A_1}, \psi_{A_2}, \tilde{\Theta}\psi_{A_1}, \tilde{\Theta}\psi_{A_2}]$. 
By introducing real hopping between the sites, we have a 3D tight-binding model. Its Bloch Hamiltonian $H^{(D)}(\bm{k})$ is given by
\begin{align}
H^{(D)}(\bm{k})= &(m+\cos k_x+\cos k_y)\sigma_0\tau_3 \notag \\
&\ +\sin k_x\sigma_3\tau_1+\sin k_y\sigma_3\tau_2+\sin k_z\sigma_3\tau_3 \notag \\
&\ \  \ +t(1+\cos k_z)\sigma_1\tau_0+t\sin k_z\sigma_2\tau_0. 
\end{align}
The tight-binding model has the  $\tilde{\Theta}$ symmetry represented by Eq.~(\ref{eq:7-2}).

The eigenenergies of $H^{(D)}(\bm{k})$ are given by
\begin{align}
    E(\bm{k}) &=\pm\sqrt{\sin^2 k_x+\sin^2 k_y+[f(\bm{k})\pm g(\bm{k})]^2},  \label{eq:D-1} 
\end{align}
where $f (\bm{k})=m+\cos k_x+\cos k_y$ and $g(\bm{k}) =\sqrt{(\sin k_z)^2+2t^2(1+\cos k_z)}$. From Eq.~(\ref{eq:D-1}), we can see the tight-binding model has two Dirac points at $\bm{k}=(0,0,0), (0,0,\pi)$ when $m=-2,\ t=0$, and has one Dirac point at $\bm{k}=(0,0,\pi)$ when $m=-2,\ t>0$. Moreover, the Dirac point at $\bm{k}=(0,0,\pi)$ has a nontrivial $Z_2$ charge $\mathcal{Q}$. This Dirac point can be transformable into a Weyl dipole by adding $\tilde{\Theta}$-preserving perturbations.
This shows that a single Dirac point (Weyl dipole) can exist in the whole BZ and it can be created or annihilated on the plane $k_z=0$.

\section{\label{sec:E}Detailed description of $H^{(A)}(\bm{k})$}
In this section, we give the detailed description of the tight-binding model in Sec.~\ref{sec:7-1}. When we put sites and define hoppings between the sites as in Sec.~\ref{sec:7-1}, the Bloch Hamiltonian
$H^{(A)}(\bm{k})$ is given by
\begin{widetext}

\begin{align}
    H^{(A)}(\bm{k}) &=H_0^{(A)}(\bm{k})+H_1^{(A)}(\bm{k})+H_2^{(A)}(\bm{k}), \\
    H_0^{(A)}(\bm{k}) &=\mqty(m_x\sigma_x+m_z\sigma_z & (t_1+t_2e^{-ik_z})I_{x}^*\sigma_0 & 0 & (t_3+t_4e^{-ik_x})I_{y}^*\tilde{I}_{z}^*\sigma_0 \\ (t_1+t_2e^{ik_z})I_{x}\sigma_0 & m_x\sigma_x-m_z\sigma_z & (t_4+t_3e^{ik_x})I_{y}^*\tilde{I}_{z}\sigma_0 & 0 \\ 0 & (t_4+t_3e^{-ik_x})I_{y}\tilde{I}_{z}^*\sigma_0 & -m_x\sigma_x-m_z\sigma_z & (t_2+t_1e^{-ik_z})I_{x}^*\sigma_0 \\ (t_3+t_4e^{ik_x})I_{y}\tilde{I}_{z}\sigma_0 & 0 & (t_2+t_1e^{ik_z})I_{x}\sigma_0 & -m_x\sigma_x+m_z\sigma_z), \\
    H_1^{(A)}(\bm{k}) &=\mqty(0 & 0 & 0 & (t_5+t_6e^{-ik_x})I_{y}^*I_{z}^* \\ 
    0 & 0 & -(t_6+t_5e^{ik_x})I_{y}^*I_{z} & 0 \\
    0 & -(t_6+t_5e^{-ik_x})I_{y}I_{z}^* & 0 & 0 \\
    (t_5+t_6e^{ik_x})I_{y}I_{z} & 0 & 0 & 0 )\sigma_3, \\
    H_2^{(A)}(\bm{k}) &=\mqty(0 & 0 & 0 &  (t_7+t_8e^{-ik_x})J_1^* \\
    0 & 0 &  (t_8+t_7e^{ik_x})J_2^* & 0 \\
    0 & (t_8+t_7e^{-ik_x})J_2 & 0 & 0 \\
     (t_7+t_8e^{ik_x})J_1 & 0 & 0 & 0 )\sigma_1,
\end{align}
\end{widetext}
where $I_{j}=1\pm e^{ik_j}\ (j=x,y,z), \tilde{I}_{z}=e^{i\phi}+e^{ik_z},\ J_1=(1+i)[1-e^{i(k_y+k_z)}]+(1-i)[e^{ik_y}-e^{ik_z}],$ and $J_2=(1-i)[1-e^{i(k_y-k_z)}]+(1+i)[e^{ik_y}-e^{ik_z}]$. 

Symmetries of the tight-binding model is as follows. Firstly, it always has the symmetry $\mathcal{G}_z\mathcal{T}=\{M_z |\frac{1}{2} 0 \frac{1}{2}\}'$ represented by
\begin{align}
    \tilde{\Theta}(\bm{k}) &=-i\mqty(0 & e^{ik_x} & 0 & 0 \\ 1 & 0 & 0 & 0 \\ 0 & 0 & 0 & e^{ik_x} \\ 0 & 0 & 1 & 0 )\sigma_1K. 
\end{align}
Moreover, when $\phi=0,\pi$, the tight-binding model also has the space-time inversion symmetry $\mathcal{PT}$ represented by
\begin{align}
    &\mathcal{PT}(\bm{k}) \notag \\
    &=i\mqty(0 & 0 & e^{-ik_y} & 0 \\ 0 & 0 & 0 & e^{-i(k_x+k_y+k_z)} \\ e^{-ik_y} & 0 & 0 & 0 \\ 0 & e^{-i(k_x+k_y+k_z)} & 0 & 0)\sigma_2K.
\end{align}

As shown in the main text, this tight-binding model has Dirac points when $\phi=0, \pi$ and pairs of Weyl points when $\phi\neq 0, \pi$ in the bulk BZ, and in both cases, DHSSs appear around the projection of the gapless nodes on the (010) surface BZ.

\section{\label{sec:F}Weak Stiefel-Whitney insulator with $\mathcal{GT}$ and $\mathcal{PT}$ symmetries}
In this section, we construct a weak SW insulator with $\mathcal{GT}$ and $\mathcal{PT}$ symmetries which do not have gapless surface states, as mentioned in Sec.~\ref{sec:5-2}. 

To this end, we consider the tight-binding model in Sec.~\ref{sec:7-2} again, and set $m=1$ and $t=0.1$. Then, one can see that the tight-binding model is gapped in the bulk, and $\omega_1=0$ and $w_2=1$ on the plane $k_y=\mathrm{const}.$. Moreover, one can also see it does not have gapless surface states on the (100) surface. Therefore, the tight-binding model is what we wanted.

\end{document}